\documentclass{JHEP3}
\usepackage{amsmath,amsfonts,amsthm,graphicx}
\usepackage{ulem}

\newcommand{\beq}{\begin{equation}}
\newcommand{\eeq}{\end{equation}}
\newcommand\beqa{\begin{eqnarray}}
\newcommand\eeqa{\end{eqnarray}}
\newcommand\bea{\begin{array}}
\newcommand\eea{\end{array}}

\newcommand{\eq}[1]{(\ref{#1})}
\newcommand{\ii}{{\bf i}}
\newcommand{\jj}{{\bf j}}
\renewcommand{\sp}{p\hspace{-1.5mm}{/}\hspace{.0mm}}
\newcommand{\sq}{q\hspace{-1.5mm}{/}\hspace{.0mm}}

\newcommand{\su}{\frak{su}}
\renewcommand{\sl}{\frak{sl}}
\newcommand{\gl}{\frak{gl}}
\newcommand{\cut}{{\circlearrowright\hspace{-3.4mm}*\hspace{0.2mm}}}

\newcommand{\mx}{x^{\rm mir}}

\newcommand{\ph}{{\rm ph}}
\newcommand{\mir}{{\rm mir}}
\newcommand{\rb}{\right)}
\newcommand{\lb}{\left(}
    \newcommand{\nn}{\nonumber}
    
    \newcommand{\COMMENT}[1]{}
    
    \newcommand{\neqa}{\nonumber\end{eqnarray}}
    \newcommand{\la}[1]{\label{#1}}

\def\a{{\alpha}}

\def\[{\left[}
\def\]{\right]}

\def\l{\lambda}
\def\e{\epsilon}

\def\s{\sigma}
\def\a{\alpha}

\def\lh{\lambda}
\def\lt{\mu}
\def\[{\left[}
\def\]{\right]}

\def\<{\langle}
\def\>{\rangle}

\def\i2{\frac{i}{2}}

\newcommand{\fm}{\scalebox{1.15}{\begin{picture}(2.8,0.0)\put(-1.2,-0.33){$\bullet$}\end{picture}}}

\newcommand{\fb}{\scalebox{0.5}{\begin{picture}(7.27,0.3)\put(-1.7,2.65){$\bigcirc$}\end{picture}}}
\newcommand{\fp}{\scalebox{0.6}[0.7]{\begin{picture}(7.24,0.2)\put(-1.5,0.5){$\bigtriangleup$}\end{picture}}}

\title{
\(\bf{\rm\bf PSU}(2,2|4)\) Character of  Quasiclassical AdS/CFT
}
\author{ Nikolay Gromov\\ DESY Theory, Hamburg, Germany \& II. Institut f\"{u}r Theoretische Physik Universit\"{a}t, Hamburg, Germany \& St.Petersburg INP, St.Petersburg, Russia \\
E-mail: \email{nikgromov$\bullet$gmail.com}}
\author{Vladimir Kazakov\footnote{member of Institut Universitaire de France}\\Ecole Normale Superieure, LPT,  75231 Paris CEDEX-5, France \&   \\
l'Universit\'e Paris-VI, France;\\
E-mail: \email{kazakov$\bullet$lpt.ens.fr}}
\author{Zengo Tsuboi\\Okayama Institute for Quantum Physics, Kyoyama 1-9-1, Okayama 700-0015, Japan \\
E-mail: \email{ztsuboi$\bullet$gmail.com}}

\abstract{
We solve the
recently proposed T- and Y-systems (Hirota equation) for the exact spectrum of AdS/CFT
in the strong coupling scaling limit for an arbitrary quasiclassical string state.
  The corresponding T-functions appear to be  super-characters of the ${\rm SU}(2,2|4)$
group in unitary representations with  a highest weight,  with the classical \({\rm AdS}_5\times {\rm S}^5\)
superstring monodromy matrix as  the group element. We propose a concise first  Weyl-type formula for these characters and show that
they correctly reproduce
the results of  quasiclassical one-loop quantization in all sectors of the superstring,
under some natural assumptions.  We also speculate about possible
relation between the T-functions and the quantum monodromy matrix.}
\preprint{OIQP-09-14\\LPT ENS-10/06 }
\usepackage{varioref}
\usepackage{amssymb}

\begin{document}

\newpage
\section{Introduction}

Recently, two of the authors and P.Vieira proposed an infinite set of equations,
the so called Y-system for AdS/CFT for the exact spectrum of
anomalous dimensions of all gauge invariant local operators in planar ${\cal N}=4$ super-Yang-Mills
theory (SYM) at arbitrary value of the  't~Hooft coupling $\lambda$ \cite{Gromov:2009tv}.
These equations were inspired by similar Y-systems in relativistic and lattice models
\cite{Zamolodchikov:1991et,Pearce:1991ty,Krichever:1996qd,Gromov:2008gj},
by  the asymptotic Bethe ansatz (ABA) \cite{Minahan:2002ve,Beisert:2005fw,Beisert:2006ez}, by the idea of the mirror theory proposed  in \cite{Ambjorn:2005wa,Arutyunov:2007tc}  with a non-relativistic single impurity dispersion relation in ${\cal N}=4$ SYM theory \cite{Santambrogio:2002sb}  and by the
structure of the leading finite size correction \cite{Bajnok:2008bm}.
In \cite{Gromov:2009bc} these equations were written in a form convenient for  numerical solution
and the anomalous dimension of the lowest lying Konishi state \cite{Bianchi:2001cm}
was found in a broad range of coupling constants $0<\lambda<700$ in \cite{Gromov:2009zb}. For the first time, the energy
of a non-protected by super-symmetry low lying state in a 4D gauge theory was found as a function of coupling
in the planar limit.
The Y-system has passed a few important checks.
The correct 4-loop result of  direct Feynman graph calculation of the Konishi anomalous dimension \(\Delta_K(\l)\)
\cite{Fiamberti:2008sm,Velizhanin:2008pc} was reproduced from the Y-system in \cite{Gromov:2009tv}\footnote{Initially the 4-loop perturbative results were reproduced in \cite{Bajnok:2008bm}
using the conjectured L\"uscher-like formulas for the world sheet theory. A similar test was successfully performed in the ABJM model - the 3-dimensional integrable analogue of the ${\cal N}=4$ SYM theory \cite{Minahan:2009wg}. These formulas presumably capture the leading finite-size corrections
and can be used up to 7-loops for the Konishi state. 5-loops where computed using this approach in \cite{Bajnok:2009vm}. Recently
the result was shown to be consistent with the Y-system approach numerically \cite{Arutyunov:2010gb}.}.
A similar  comparison for the length 3 operators at 5-loops also confirms
the validity of the Y-system \cite{Fiamberti:2009jw}.
The extrapolation of  numerical results of \cite{Gromov:2009zb} to the strong coupling
 was found to be approximately $\Delta_K\simeq 2.0004\lambda^{1/4}+1.99/\lambda^{1/4}+\dots$
 where the leading coefficient agrees with the string prediction $2\sqrt{n}\;\lambda^{1/4}$ for $n=1$ \cite{Gubser:1998bc}\footnote{
 The subleading terms are still a challenge for  string theorists. Two
different values for $1/\lambda^{1/4}$ coefficient
 were obtained on the string side \cite{Roiban:2009aa,Arutyunov:2005hd} on
 the basis of rather bold assumptions.
 The results \cite{Arutyunov:2005hd} are obtained for a truncated
model  where quantum contribution of some string modes are ignored
 whereas in \cite{Roiban:2009aa} the applicability of the quasi-classics
in the small charge limit was assumed.
It is also important to mention that the prediction of \cite{Gromov:2009zb}  for that coefficient was subject to the
explicit assumption that the first expansion coefficients are of order of $1$
and do not grow too rapidly so that the numerical results for $\lambda\lesssim 700$ allow for this prediction.
In \cite{KZ} a warning was rased that this assumption could be not  correct.
 Hopefully the direct world sheet computation
 in the Metsaev-Tseytlin superstring sigma model \cite{Metsaev:1998it}
 along the lines  of  other approaches mentioned in
\cite{Roiban:2009aa} will be done soon and will lift this uncertainty.
 }.
Also the  general  asymptotic solution  of the Y-system  for long operators was constructed \cite{Gromov:2009tv}
and its consistency with  the asymptotic Bethe ansatz (ABA) of \cite{Beisert:2005fw,Beisert:2006ez}
was shown.
As we will see this asymptotic solution plays a fundamental role in the whole construction since it
defines the analytic and asymptotic properties
of the exact solution   of the Y-system for a given physical state at a finite $L$.
In particular it was used in \cite{Gromov:2009bc}
to write the integral equation for  excited states.
Another important test was done in \cite{Bombardelli:2009ns,Gromov:2009bc,Arutyunov:2009ur}
where the Y-system was obtained from the Al.~Zamolodchikov thermodynamic Bethe ansatz (TBA) approach for the BMN ground state\footnote{The ground state
by itself is protected and has zero anomalous dimension. Nevertheless the TBA equations capture
some important structural information about the Y-system.}.
\begin{figure}[ht]
\begin{center}
\includegraphics[scale=0.7]{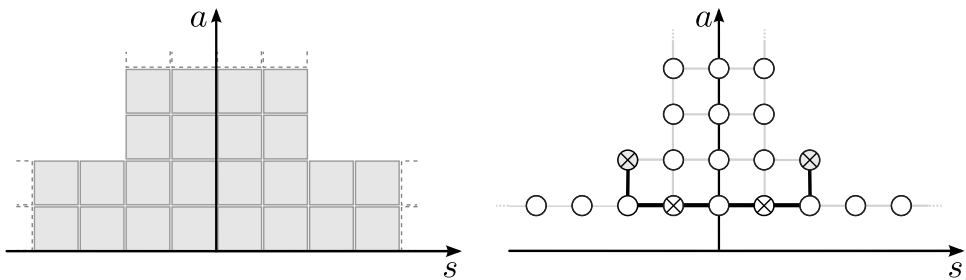}
\end{center}
\caption{
(Left:\ Fig.1a; Right:\ Fig.1b)
 \textbf{T}-shaped ``fat hook" (\textbf{T}-hook) uniting two ${\rm SU}(2|2)$ fat hooks,
see \cite{Gromov:2009tv} for this \textbf{T}-hook and its generalization \cite{Hegedus:2009ky}.
}\label{T-Hook}
\end{figure}

Whereas the classical finite gap solutions of Metsaev-Tseytlin sigma model \cite{Kazakov:2004qf,Beisert:2005bm}
stemming from the world sheet integrability \cite{Bena:2003wd} and calculating the
dimensions of long operators in a strong coupling regime follow from the
ABA equations
\cite{Beisert:2006ez} in a direct and simple way, the quasiclassical one-loop corrections,
also available for an arbitrary finite gap  solution \cite{Frolov:2002av,Gromov:2009zza},
are already in a severe disagreement with the ABA even at infinite length in the scaling $L\sim\sqrt\lambda\to\infty$ \cite{SchaferNameki:2006ey}\footnote{
The ABA agrees with the one-loop corrections when
$L/\sqrt\lambda\gg 1$ \cite{SchaferNameki:2005tn,FS1,Gromov:2007ky}.}.
Recently it was shown \cite{Gromov:2009tq} that the Y-system cures this one-loop disagreement
generically for any classical states in $AdS_3\times S^1$ and thus takes into account infinitely many
finite size wrapping contributions\footnote{By wrapping contributions we mean essentially all the finite size corrections to the ABA. Historically the term refers to specific Feynman graphs of SYM running around all the legs of a local operator under study.}.
Similar results were also obtained for several subsectors of the ABJM theory
\cite{ABJM,Gromov:2009at} where the Y-system was also conjectured \cite{Gromov:2009tv,Bombardelli:2009xz,Gromov:2009at}.
These results deeply test  the structure of the Y-systems since all
wrapping contributions are crucially important in that case.

In this paper we construct the complete solution of the AdS/CFT Y-system in the
strong coupling limit generalizing \cite{Gromov:2009tq}
and reproduce from this solution the equations arising for the
quasiclassical one-loop corrections of \cite{Gromov:2009zza}.
In this limit, the finite difference operators w.r.t the spectral parameter disappear from   the Y-system and it reduces to a simplified system
of equations called in the mathematical literature the Q-system:\footnote{A terminological note: One should not confuse the Q-system with the Baxter \(Q\)-operators. Q-system is a simplified version of  Hirota equation \vref{Tsystem}.  We often call  the last one as T-system. Baxter equations can be called TQ-relations and the particle-hole duality relations among supersymmetric Baxter functions are called QQ-relations.}
\(
T_{a,s}^2 =T_{a+1,s}T_{a-1,s}+T_{a,s+1}T_{a,s-1}\;.
\)

This equation is related to the Y-system in the considered limit by $Y_{a,s}=\frac{T_{a,s+1}T_{a,s-1}}{T_{a+1,s}T_{a-1,s}}$. The Q-systems are frequently used as the  defining
equations for the characters of representations of the underlying symmetry groups
(see for example \cite{Q-systems,Tsuboi:1997iq,Kazakov:2007fy}). Here we constructed the general solution of such a
Q-system with the T-hook boundary conditions with respect to the representational
indices \(a,s\) (see Fig.\ref{T-hook2}).
We will argue that the solution is given in terms of characters of certain
unitary representations of \({\rm SU}(2,2|4)\) group.
It demonstrates in a nontrivial way that the full global symmetry  of AdS/CFT is present
in the Y-system in spite of the original \({\rm SU}(2|2)\times {\rm SU}(2|2)\) setting due to the
choice of the light cone gauge. Then, using the asymptotic solution,
we fix the parameters of the general solution. This leads  to a very simple result
\beq
T_{a,s}={\rm Str}_{a,s}\Omega(x)\;,
\eeq
where $\Omega(x)$ is the classical monodromy matrix \cite{Bena:2003wd} and the trace is taken in some
particular representations labeled by $a,s$. We also speculate that a similar relation
should hold at the quantum level.
Let us also note that the quantum generalization of this character solution
of the AdS/CFT Y-system  would reduce the problem at any coupling to a finite
set of non-linear integral equations similar to Destri-DeVega equations known
for some relativistic sigma models. This should be possible due to the underlying
integrable discrete Hirota dynamics of the Y-system  \cite{Gromov:2008gj}
(see \cite{Tsuboi:2009ud,Hegedus:2009ky} for the first steps).
The general strong coupling solution we present here can be an important
step on this way.

In Sec.\ref{sec:2old}  we present the general  character  solution of the Q-system
(or of the related Y-system), for the T-hook
boundary conditions  reflecting  the global ${\rm SU}(2,2|4)$ symmetry of AdS/CFT
problem. At the end of Sec.\ref{sec:2old} these characters will be presented in a new, explicit and concise  form in terms of determinants of $2\times 2$ and $4\times 4$ matrices reminding the 1-st Weyl formula;
a detailed description of these representations
is given in the Sec.\ref{AppRep}.   In Sec.\ref{sec:3old} we compare this solution of Y-system and Q-system
  to the quasiclassical (one-loop) spectrum of the theory. The Sec.\ref{sec:4old}
  will summarize our results and propose some new possible directions in testing and simplifying  the AdS/CFT Y-system.

\section{Y-system of AdS/CFT and the characters of ${\rm U}(2,2|4)$}\la{sec:2old}
In this section we will remind the formulation of the AdS/CFT Y-system and construct the complete solution  in an important particular case: when the dependence on the spectral parameter
is slow and the finite shifts can be neglected. In this case, the Y-system, or the equivalent T-system (Hirota bilinear difference equation),
 does not contain the shifts in \(u\)
and is usually called the Q-system.  The Q-system, a finite difference equation with respect to a
couple of discrete variables, can be interpreted as an equation for characters of particular
irreducible representations (usually with
\(a\times s\) rectangular Young diagrams, with \(a\) and \(s\) being its size in the antisymmetric
and symmetric directions, respectively) for a given symmetry group. A specific group enters the
Q-system only through  the  boundary conditions w.r.t. the discrete variables parameterizing
the representation space. We will give in this section the full solution of such a
Q-system with the T-hook boundary conditions (see Fig.\ref{T-Hook}a), relevant to the AdS/CFT Y-system \cite{Gromov:2009tv},
in an explicit and concise form and interpret them as super-characters of some unitary representations of the
\({\rm U}(2,2|4) \) group.
 \subsection{Y-system for AdS/CFT: equations and definitions }
 Y-system encoding the spectrum of all local operators  in planar AdS/CFT correspondence
 \cite{Gromov:2009tv} is a set of functional equations\footnote{
We shall always denote \(f^\pm=f(u\pm i/2)\) or even more generally \(f^{[\pm a]}=f(u\pm a\,i/2)\).
In this equation we choose the Y-functions to have branch cuts going to infinity.
The analytic properties of the Y-functions
could be seen from the asymptotic solution described below.}
\begin{equation}
\label{eq:Ysystem} \frac{Y_{a,s}^+ Y_{a,s}^-}{Y_{a+1,s}Y_{a-1,s}}
 =\frac{(1+Y_{a,s+1})(1+Y_{a,s-1})}{(1+Y_{a+1,s})(1+Y_{a-1,s})} \,.
\end{equation}
The functions \(Y_{a,s}(u)\) are defined only on the nodes marked by gray and white circles  in Fig.\ref{T-Hook}b.
Solutions
of Y-system with appropriate analytic properties
define the energy of a state (anomalous dimension of an operator in ${\cal N}=4$ SYM) through the formula
\begin{equation}
E=\sum_{j}\epsilon^{\ph}_1(u_{4,j})+\sum_{a=1}^\infty\int_{-\infty}^{\infty}\frac{du}{2\pi i}\,\,\frac{\partial\epsilon_a}{\partial u}\log\left(1+Y_{a,0}(u)\right) , \label{eq:Energy}
\end{equation}
where the physical dispersion relation between the energy \(\e_a\) and the momentum $p_a$ is  parameterized in terms of the rapidity (spectral parameter) \(u\) as follows
\begin{equation}\label{eq:energyU}
\epsilon_a(u)= a+\frac{2ig}{x^{[+a]}}-\frac{2ig}{x^{[-a]}}
\end{equation}
where $g=\frac{\sqrt\lambda}{4\pi}$. We consider two different branches of the double valued function $x(u)$
\beq\la{xxs}
\qquad x^\ph(u)=\frac{1}{2}\lb \frac{u}{g}+\sqrt{\frac{u}{g}-2}\;\sqrt{\frac{u}{g}+2} \rb
\;\;,\;\;\mx(u)=\frac{1}{2}\lb \frac{u}{g}+i\sqrt{4-\frac{u^2}{g^2}}\rb \,.
\eeq
They coincide above the real axis. $x^\ph(u)$ is defined to have a finite branch cut between $\pm 2g$
whereas in $x^\mir(u)$ the cut is chosen to go through infinity along the real axes.
If it is not stated otherwise we always define $x(u)=x^\mir(u)$.
The
rapidities $u_{4,j}$ are fixed by the exact  Bethe ansatz equations
for any size \(L\) of the SYM operators
\beq
Y^\ph_{1,0}(u_{4,j})=-1\,.
\eeq
The Y-system for AdS/CFT can be rewritten as  Hirota bilinear difference equation (T-system)
\cite{Gromov:2009tv}\footnote{In a sense the T-system is more fundamental
than the  Y-system.
Any explicit solution of Y-system looks simpler in terms of T's. Moreover for the T-hook,  two equations, for $(a,s)=(2,2)$ and $(2,-2)$, are missing
in  the Y-system. One cannot write the  equations for these nodes
in terms of Y-functions in a ``local" functional form. However, these equations are present in a local form in T-system.
We will clearly see this while solving the T-system for strong coupling.}
\begin{equation}
 \label{Tsystem}
 T_{a,s}^+T_{a,s}^- =T_{a+1,s}T_{a-1,s}+T_{a,s+1}T_{a,s-1} \,, \\
\end{equation}
where
\beq
Y_{a,s}=\frac{T_{a,s+1}T_{a,s-1}}{T_{a+1,s}T_{a-1,s}}\;.
\eeq
The functions \(T_{a,s}(u)\) are non-zero only on the 2D lattice drawn on Fig.\ref{T-Hook}a.
\beq
T_{a,s}=0\;\;{\rm if}\;\;a<0\;\; {\rm or}\;\; a,|s| >2\;.\la{boundary}
\eeq
There is a ``gauge" freedom in \eq{Tsystem}
\beq
T_{a,s}(u)\to G_1\lb u+i\frac{a+s}{2}\rb
G_2\lb u+i\frac{a-s}{2}\rb
G_3\lb u-i\frac{a+s}{2}\rb
G_4\lb u-i\frac{a-s}{2}\rb T_{a,s}(u)
\eeq
which maps one solution of the T-system to another but leaves the $Y_{a,s}$ intact.

\subsection{The   Q-system  limit and its ${\rm U}(2,2|4)$  ``character" solution}
In the strong coupling limit \(\lambda\to\infty\)   which we  shall study in this paper,
we notice that the \(\lambda\)-dependence can be scaled out from the formulas like
\eqref{xxs} by simple rescaling \(u\to 2g z\) with $g=\frac{\sqrt\lambda}{4\pi}$. Then the \(u\)-shifts in   the
Y-system \eqref{eq:Ysystem} and T-system \eqref{Tsystem}    become negligible and it can be now written  as a
Q-system\footnote{The shifts in the spectral parameter cannot be neglected close to the branch cuts of $T_{a,s}$ going along the real axis with $|u|>2g$.}
\begin{align}
T_{a,s}^2 =T_{a+1,s}T_{a-1,s}+T_{a,s+1}T_{a,s-1} \,,\label{Hirota}
\end{align}
with the AdS/CFT T-hook boundary conditions \eq{boundary}.
The gauge transformations for the  the Q-system are reduced to
\beq
T_{a,s}\to g_1\; g_2^a\; g_3^s\; g_4^{as}\;T_{a,s}\,\,.
\eeq

\begin{figure}[t]
\begin{center}
\includegraphics[scale=0.5]{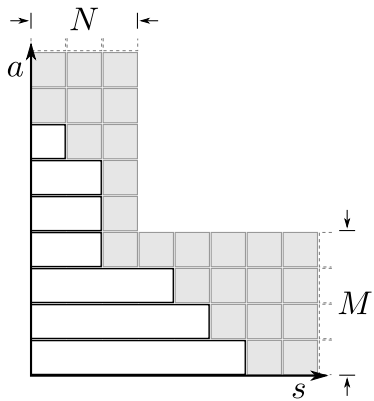}
\end{center}
\caption{ ``Fat hook" where  the representations of a \(SU(M|N)\) symmetric super-spin chain live.   See \cite{Tsuboi:1997iq,Kazakov:2007fy} for the details on fat hooks and
$T$-functions for spin chains related to superalgebras.}\label{FatHook}
\end{figure}
We shall now construct the general solution of this  Q-system in terms of certain ``characters" of
${\rm GL}(4|4)$   group. We will show in the next
 section that these characters lead, after an appropriate identification of their parameters
 with  the quasi-momenta of the classical monodromy matrix, to exactly the same expressions as in the
 full quasiclassical (one-loop)
solution of  MT string sigma-model \cite{Gromov:2007ky} thus generalizing  the results of \cite{Gromov:2009tq}
for the ${\rm SL}(2)$ subsector.
The representations corresponding to the solution in T-hook
are certain infinite dimensional representations described in Sec.\ref{AppRep}.
We first start from a more standard example of  finite dimensional representations of   ${\rm GL}(4|4)$   group.
The super characters for ``symmetric" representations \(T_{1,s}\), or supersymmetric Schur polynomials, for the group \({\rm GL}(M|N)\) are defined through the following generating function
(see e.g.  \cite{Conrey05} which will be relevant to our discussion):
\begin{equation}\label{KMgen}
w_{M|N}(t;g)={\rm Sdet}\frac{1}{1-g\, t}=\frac{\prod_{n=1}^N(1-y_n\,t)}{\prod_{m=1}^M(1-x_m\,t)}=\sum_{s=1}^\infty t^s\,T^{(M|N)}_{1,s}[g]
\;,
\end{equation}
where $(x_1,\dots,x_M|y_1,\dots,y_N)$ are the eigenvalues of a group element $g\in {\rm GL}(M|N)$\footnote{We consider the generic diagonalized group element}.
The rest of the characters for ``rectangular" representations (for which the Young diagram contains \(a\) columns and \(s\)-rows) can be calculated using the Jacobi-Trudi  formula:
\begin{equation}\label{eq:Jacobi-Trudi}
T_{a,s}=\det_{1\le i,j\le a}\, T_{1,s+i-j}\;.
\end{equation}
These characters are non-zero only in a ``fat hook",
or $[M|N]$-hook presented on the Fig.\ref{FatHook}. The generating function of ${\rm GL}(4|4)$ characters can  be represented as
\begin{eqnarray}
w_{4|4}(t;g)
=\frac{(1-y_{1}t)(1-y_{2}t)}
{(1-x_{1}t)(1-x_{2}t)}\times\frac{(1-y_{3}t)(1-y_{4}t)}
{(1-x_{3}t)(1-x_{4}t)}
\label{GEN44}
\end{eqnarray}
i.e.,
\begin{equation}
T_{1,s}^{(4|4)}[g]=\oint_{C_0}\frac{dt\,\, w_{4|4}(t;g) }{2\pi i}t^{-s-1}\;.
\label{oint44}\end{equation}
Notice that the integrand has also poles at $t=1/x_j$ in addition to the pole at the origin.
To get \eq{KMgen}
one should encircles only the point \(t=0\) and leave outside all other poles.
Using that $w_{4|4}(t;g^L\otimes g^R)=w_{2|2}(t;g^L)\times w_{2|2}(t;g^R)$,
where $g^L,g^R\in {\rm GL}(2|2)$
we can represent $T_{1,s}^{4|4}$ in a specific  form
\beq
T_{1,s}^{(4|4)}[g^L\otimes g^R]=
 \sum_{j=0}^{s}T^{(2|2)}_{1,s-j}[g^L]\times T^{(2|2)}_{1,j}[g^R]\;.
\eeq
What would be the analog of these characters satisfying the Q-system \eqref{Hirota}
 and the T-hook boundary conditions \eqref{boundary}?
 A natural definition appears to be the same eq.\eqref{oint44}, after a simple change of the integration contour: we encircle this time \(t=0\) together with the  poles \(\frac{1}{x_3},\frac{1}{x_4}\,\)
 corresponding to the second  subgroup \({\rm GL}(2|2)\),     leaving outside
 the poles \(\frac{1}{x_1},\frac{1}{x_2}\,\) corresponding to the first subgroup  \({\rm GL}(2|2)\).
 This amounts  to expanding the first factor in \eqref{GEN44} in \(t\) and the second one in \(1/t\) and picking the power \(t^{s}\):
\begin{align}
&T_{1,s}[g^L\otimes g^R]=
\frac{y_{3}y_{4}}{x_{3}x_{4}}\sum_{j=\max(0,-s)}^{\infty}
T^{(2|2)}_{1,s+j}[g^L]\times T^{(2|2)}_{1,j}[1/g^R].
\label{oint224}\end{align}
Note that unlike the finite-dimensional representations \eqref{oint44}, the r.h.s
of \eqref{oint224} contains infinite number of terms and thus
such  characters correspond to some infinite dimensional unitary representations
of the group \({\rm U}(2,2|4)\). Then one can see that the Jacobi-Trudi
formula \eq{eq:Jacobi-Trudi}
gives the full solution of the  Q-system for the T-hook Fig.\ref{T-Hook}a
with the boundary conditions \eq{boundary}!

Let us mention here that this method of construction of characters, or even quantum transfer matrices of  infinite dimensional representations of \(u(2,2|4)\)  was already mentioned in \cite{Beisert:2005di} where it was applied to the AdS/CFT one-loop Bethe ansatz. It was proposed there (see Appendix B) to expand some of the monomials of the generating functional (the analogue of our  \(w_{4|4}(t;g);\) proposed by Krichever et al. in \cite{Krichever:1996qd} and generalized to the supersymmetric case
in  \cite{Tsuboi:1997iq})  in  positive powers, and some - in negative powers  of the shift operator \(D=e^{-i\partial_u/2}\). Note that this method is enough to generate the general solution of the full quantum Hirota equation \eqref{Tsystem} within the \((2,2|4)\) \(\mathbb{T}\)-hook by means of the
 Bazhanov-Reshetikhin type formula
\begin{equation}\label{eq:BR}
T_{a,s}=\det_{1\le j,k\le a}T_{1,s+j-k}\left(u-\i2(a-k-j+1)\right),
\end{equation}
where the T-functions \(T_{1,s}(u)\) are generated by the following generating functional
(the direct generalization of \eqref{GEN44}
\begin{multline}\nn
W=\[(1-DY_1 D)
\frac{1}{1-DX_1 D}
\frac{1}{1-DX_2 D}
(1-DY_2D)\]_+\\
 \times \[(1-DY_{3}^{}D^{})
\frac{1}{1-D^{}X_{3}^{}D^{}}\frac{1}{1-D^{}X_{4}^{}D}
(1-DY_{4}D)\]_-
=\sum_{s=-\infty}^{\infty} D^{s}T_{1,s}D^{s} .
\end{multline}
 Here \(\{Y_1(u)|X_1(u),X_2(u)|Y_2(u),Y_3(u)|X_3(u),X_4(u)|Y_4(u)\}\) are 8 arbitrary functions of the spectral \(u\) parameterizing the general solution.
One expands here in positive powers of the operator  \(D\)
 inside the bracket \(\left[\dots\right]_+\) corresponding to the \(u(2|2)_R\) subalgebra, and in negative powers of \(D\) inside the bracket \(\left[\dots\right]_-\) corresponding to the \(u(2|2)_L\) subalgebra.
 We hope to explore this general solution in  the future for the construction of finite system of non-linear integral equations for the AdS/CFT spectrum.
\footnote{ A solution in the \(\mathbb{T}\)-hook was already constructed in \cite{Hegedus:2009ky} on the base of the B\"acklund transformations of \cite{Kazakov:2007fy}.}

  In the next section, we  will write the above $T_{a,s}$    for any \(a,s  \) in a   concise and explicit form, similar to the first Weyl formula for characters,
 through  Wronskian-like determinant expressions of certain \(2\times 2\) and \(4\times 4\) matrices.

\subsection{New determinant formulae}
The sum in \eqref{oint224} can be calculated explicitly and the result can be presented in the following remarkable determinant form:
\beqa\label{CharSol}
T_{a,s}=
\left\{\bea{cc}
(-1)^{(a+1)s}\left(\frac{x_3x_4}{y_1y_2y_3y_4}\right)^{s-a}\frac{{\rm det}\left(S_i^{\theta_{j,s+2}} y_i^{j-4-(a+2)\theta_{j,s+2}}\right)_{1\le i,j\le 4}}
{
{\rm det}\left(S_i^{\theta_{j,0+2}} y_i^{j-4-(0+2)\theta_{j,0+2}}\right)_{1\le i,j\le 4}
}&,\;\;a\geq|s|\\
\frac{{\rm det}\left(Z_i^{(1-\theta_{j,a})} x_i^{2-j+(s-2)(1-\theta_{j,a})}\right)_{1\le i,j\le 2}}
{
{\rm det}\left(Z_i^{(1-\theta_{j,0})} x_i^{2-j+(0-2)(1-\theta_{j,0})}\right)_{1\le i,j\le 2}
}
&,\;\;s\geq +a
\eea\right.
\eeqa
where
\beqa
S_i&=&\frac{(y_i-x_3)(y_i-x_4)}{(y_i-x_1)(y_i-x_2)}\\
Z_i&=&\frac{(x_i-y_1)(x_i-y_2)(x_i-y_3)(x_i-y_4)}{(x_i-x_3)(x_i-x_4)}\;,
\eeqa
and
\beqa
\theta_{j,s}=\left\{\bea{cc}
1&,\;\;j>s\\
0&,\;\;j\le s
\eea\right.\;.
\eeqa
The other $T$'s can be obtained using the wing-exchange symmetry
which is related to an outer automorphism of the Dynkin diagram of $\gl(4|4)$
\beq
T_{a,s}(x_{1},\dots,x_{4}|y_{1},\dots,y_{4}) =
\lb\frac{y_1y_2y_3y_4}{x_1x_2x_3x_4}\rb^a T_{a,-s}\left(\left.\frac{1}{x_{4}},\dots,\frac{1}{x_{1}}\right|\frac{1}{y_{4}},
\dots,\frac{1}{y_{1}}\right).
\eeq
Note that for the $\sl(4|4)$ case the first factor in the r.h.s. is absent.

These formulae are summarized in the Appendix \ref{AppMath} in the \textit{{Mathematica}} form.
Note that the upper part of the T-hook is represented by a $4\times 4$ determinant reminding the 1-st
Weyl formula for ${\rm GL}(4)$ characters
 (it would be them if all $S_i$ were equal to $1$). The left and right wings are presented by $2\times 2$
  determinants similar to ${\rm GL}(2)$ characters.
Hence, we can identify the variables $y_1,y_2,y_3,y_4$ as the eigenvalues from ${\rm U}(4)$ subgroup
 of ${\rm SU}(2,2|4)$, the variables $x_1,x_2$ as the eigenvalues of the ${\rm U}_R(2)$ subgroup and $x_3,x_4$ as
  the eigenvalues of
 the ${\rm U}_L(2)$ subgroup.
Our solution of the Q-system is symmetric under any permutations of $y_i$
and the permutations of $x_1\leftrightarrow x_2$ and $x_3\leftrightarrow x_4$.
However the solution is not invariant under the full Weyl group of ${\frak{gl}(4|4)}$
which includes arbitrary permutations of $x_i$.
The origin of this ``symmetry breaking'' will go back to the fact that
$T_{a,s}$ is
a super-character of an {\it infinite} dimensional representation.
Another important property, under the rescaling of eigenvalues,
reads
\beq
T_{a,s}(\alpha x_a,\alpha y_a)=\alpha^{a s} T_{a,s}(x_a,y_a)\;,\la{scT}
\eeq
which means that $T_{a,s}$  is a homogeneous function of degree $as$.

\section{Description of  representations of  ${\rm U}(2,2|4)$ for the AdS/CFT Y-system }\la{AppRep}
In this section we describe in details the representations of
${\rm U}(2,2|4)$ group corresponding to the super-characters
from the previous section.

\begin{figure}[th]
\begin{center}
\includegraphics[scale=0.4]{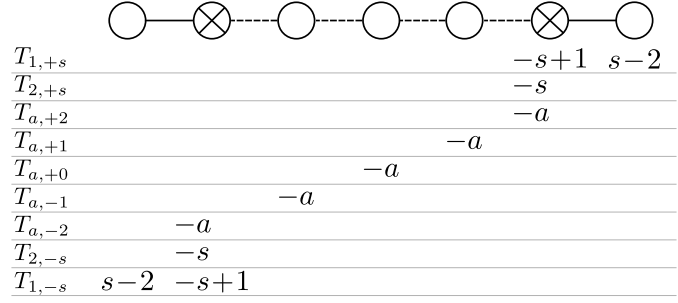}
\end{center}
  \caption{Dynkin diagram for the Lie superalgebra
  $\frak{gl}(4|4)$ with the Dynkin indexes corresponding to the characters in (2.18).
}
\label{DynkinL}
\end{figure}

\subsection{Description of the general construction}
In \eqref{CharSol}
we presented a formal solution of  Hirota (Q-system)
equations with the T-hook boundary conditions.
At the same time it is known that the Q-systems with various boundary conditions
can be often solved by characters of representations with rectangular Young diagrams.
In this Section, we describe a class of representations of  the superalgebra ${\gl}(4|4)$
leading to the super-character solution \eq{CharSol}.\footnote{V.K. thanks N.Beisert for inspiring discussions on this subject.}
Provided the Y-system is indeed a set of  equations encoding the
exact AdS/CFT spectrum of the theory, these representations
 clearly have some physical importance.
These representations   live in the mirror  space of the theory and reflect the \({\rm SU}(2,2|4)   \) symmetry properties of the mirror ``particles". In analogy to the spin chain terminology, we would call the
mirror space as auxiliary space, in contradistinction to the physical space where the physical ``particles" live.

Comparing our formula \eqref{CharSol} with similar
formulas for characters of the refs.\cite{CLZ03,Kwon06} we find that we
deal  with
 a class of representations called
``unitarizable irreducible $\frak{gl}(M_1|N|M_2)$-modules''\footnote{or $\frak{gl}_{M_1+M_2|0+N}$ in
the notations of \cite{CLZ03}}.
We denote them as  $W(M_1|N|M_2;\lambda)$.
They are infinite dimensional irreducible highest weight representations of $\frak{gl}(M|N)$ where $M=M_1+M_2$
with respect to a (non-standard) Borel subalgebra corresponding to the $(M_1|N|M_2)$ grading described below.
They are the unitary representations of ${\frak u}(2,2|4)$.

The algebra is generated by standard super-generators
$E_{ab}$. In  fundamental representation $(E_{ab})_{ij}=\delta_{ia}\delta_{jb}$ and they obey
the following super-commutation relation
\beq
[E_{ab},E_{cd}\}=
\delta_{bc}E_{ad}-(-1)^{(p_a+p_b)(p_c+p_d)}\delta_{da}E_{cb},
\eeq
where the grading $p_i=1$ for
$i=M_{1}+1,M_{1}+2,\dots,M_{1}+N$ and is $0$ otherwise.

We define the Cartan subalgebra ${\mathfrak h}$
and Borel subalgebra ${\mathfrak b}$  as follows
\beq
{\mathfrak h}=\sum_{a =1}^{N+M}{\mathbb C}E_{aa}\;\;,\qquad \qquad
{\mathfrak b}= \sum_{a \le b}{\mathbb C}E_{ab}\; .
\eeq
Note that for the definition of the Borel subalgebra the ordering of  indexes is crucial
and we chose $M_1$ bosonic components, followed by $N$ fermionic and then again by $M_2$ bosonic.

Then we introduce the  space ${\mathfrak h}^{*}$
dual to the Cartan subalgebra ${\mathfrak h}$. Let  $\varepsilon_i$
be a graded basis of the dual space ${\mathfrak h}^{*}$ of the Cartan subalgebra ${\mathfrak h}$
such that $\varepsilon_{i}(E_{jj})=\delta_{ij}$.
We define a bilinear form $(\cdot|\cdot)$ in ${\mathfrak h}^{*}$:
\beq
(\varepsilon_{i}|\varepsilon_{j})=(-1)^{p_i}\delta_{ij}\;.
\eeq
The simple root system in this basis is given as follows:
\beq\la{alphai}
\alpha_i=\varepsilon_i-\varepsilon_{i+1}\;\;,\;\;i=1,2,\dots,N+M-1\;.
\eeq
\begin{figure}[t]
\begin{center}
\includegraphics[scale=0.5]{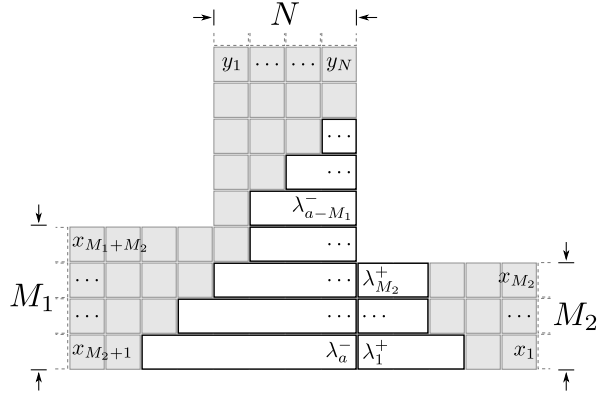}
\end{center}
\caption{ ``T- hook"and the highest weight components arranged as a generalized Young diagram living inside the T-hook
}\label{T-hook2}
\end{figure}
The class of  representations we would like to describe is parameterized by
a generalized partition $\lambda=(\lambda_{1},\lambda_{2},\dots, \lambda_{a})$
where   $\lambda_{1},\lambda_{2},\dots, \lambda_{a} \in {\mathbb Z}$ (not necessarily positive)
and ordered
$\lambda_{1} \ge \lambda_{2} \ge \dots \ge \lambda_{a}$ (see Fig.\ref{T-hook2}).
Let $\lambda $ be a generalized partition such that
$-N \le \lambda_{a-M_1}$, $\lambda_{M_2+1} \le 0$.
Then
$W(M_1|N|M_2;\lambda)$ is defined as the representation with  the
 highest weight
\begin{align}\la{weight}
\Lambda =-\sum_{i=1}^{M_1}(\langle \lambda^{-}_{i+a-M_1}-N \rangle +a)\varepsilon_{i}-
\!\!\!\!
\sum_{i=M_1+1}^{M_1+N}((\lambda^{-})^{\prime}_{N+M_1+1-i} -a)\varepsilon_{i}
+\sum_{i=N+M_1+1}^{N+M} \lambda^{+}_{i-N-M_1}\varepsilon_{i},
\end{align}
where we introduced the symbols
\begin{gather}
\langle x \rangle :=\max (x,0)\;\;,\;\;
\lambda^{+}:=(\langle \lambda_{1} \rangle,\langle \lambda_{2} \rangle,\dots,\langle \lambda_{a} \rangle)\;\;,\;\;
\lambda^{-}:=(\langle -\lambda_{1} \rangle,\langle -\lambda_{2} \rangle,\dots,\langle -\lambda_{a} \rangle)\;,
 \nonumber
\end{gather}
and by \((\l^{\pm})'\) we denote a conjugate partition obtained from the usual partition \(\l^{\pm}\),
with only positive entries, by the reflection of the associated  Young diagram w.r.t. its main diagonal\footnote{formally defined as
$(\lambda^{\pm})^{\prime}_{j}= {\rm Card}\{ k | (\l^{\pm})_{k} \ge j \}$.
}.

The character formulae of these representations are given in \cite{CLZ03}.
In our case we have to take $M_1=M_2=2,\;N=4$ and the generalized partition
is represented by a rectangular $a\times (s-2)$ Young diagram
\begin{align}\label{lasa}
\lambda=(\underbrace{s-2,\dots,s-2}_a).
\end{align}
On the level of representation,
the Q-system for these representations follows from the decomposition of
the tensor product of representations in the  Theorem 6.1 in \cite{CLZ03}.

Now let us consider the case $\frak{gl}(2|4|2)$ which is of the prime importance for us.

For the rectangular diagram \eq{lasa} the weight  \eq{weight} is
\beqa\la{weight242}
\Lambda =
\left\{
\bea{llll}
a\Big(-\varepsilon_1-\varepsilon_2\Big)&\!+\;(s+2)\sum\limits_{i=7}^{a+6} \varepsilon_{i}
&,\;\; \;\;\;s<-2&,\;\; 0\le a\le 2\\
a\Big(-\varepsilon_1-\varepsilon_2+\sum\limits_{i=3}^{s+4}\varepsilon_{i}\Big) &&,\;\;-2\le s\le 2&,\;\;0\le a\\
a\Big(-\varepsilon_1-\varepsilon_2+\sum\limits_{i=3}^{6}\varepsilon_{i}\Big)&\!+\;(s-2)\sum\limits_{i=7}^{a+6}\varepsilon_i&,\;\;\;\;\; 2<s&,\;\; 0\le a\le 2
\eea
\right.\;.
\eeqa
The  Kac-Dynkin labels\footnote{
Here we define the Kac-Dynkin labels as \(b_{j}=(\lambda | \a_{j} )\). In the mathematical literature
it is usually normalized as \(b_{j}=2(\lambda | \a_{j} )/(\a_{j} | \a_{j} ) \)
 for \( (\a_{j} | \a_{j} ) \ne 0 \).
}
can be easily calculated (see Fig.\ref{DynkinL}).
The parameters $(x_1,\dots,x_4)$ and $(y_1,\dots,y_4)$ entering  \eq{CharSol}
can be defined in our notations as  formal exponentials
\begin{align}\la{xyident}
x_{3}=e^{\varepsilon_1}\;, \;
x_{4}=e^{\varepsilon_2}\;| \;
y_{1}=e^{\varepsilon_3}\;, \;
y_{2}=e^{\varepsilon_4}\;, \;
y_{3}=e^{\varepsilon_5}\;, \;
y_{4}=e^{\varepsilon_6}\;|\;
x_{1}=e^{\varepsilon_7}\;, \;
x_{2}=e^{\varepsilon_8}\;.
\end{align}
The way we identify $x_i$ with \(\varepsilon_i\)'s is somewhat  nontrivial.
This notation means that for example $x_3$ for a given
element $h$ of the Cartan subalgebra returns the first eigenvalue
$x_3(h)$ of the corresponding group element in the fundamental $8$ dimensional representation\footnote{
In \cite{CLZ03}, the authors consider characters, while we are dealing with supercharacters. Thus
one has to change the sign of $y_{i}$ to compare our formulae with the
character formulae in \cite{CLZ03}.}.

\begin{figure}[th]
\begin{center}
\includegraphics[scale=0.4]{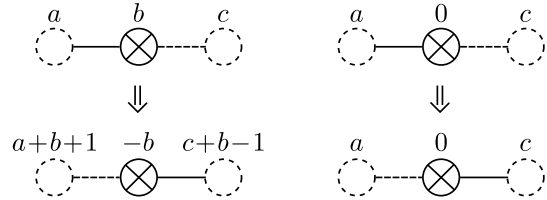}
\end{center}
\caption{Transformation property of the Dynkin labels under the fermionic duality. The duality
transform the diagram in one grading to another. The dotted lines correspond to the fermionic grading
whereas the solid lines represent bosonic grading.}\label{duality}
\end{figure}

One may want  to transform the Dynkin labels to  different gradings.
For that one can use the Weyl reflection with respect to the odd simple roots \cite{LSS} (see Fig.\ref{duality}).

 Q-system is a set of functional relations among characters of representations of
Yangians, or quantum affine algebras.
 Thus a solution of the  Q-system  is in general a linear combination of characters of
  the Lie algebra. But for the super Yangian $Y(\frak{gl}(M|N))$, it is just
 a super-character of $\frak{gl}(M|N)$ since the
evaluation map from $Y(\frak{gl}(M|N))$ to $\frak{gl}(M|N)$ allows one to
 lift the representations of $\frak{gl}(M|N)$
to those of $Y(\frak{gl}(M|N))$.
We find that this is also the case with our AdS/CFT Q-system.

\subsection{Unitarity}
As we mentioned, the class of  representations described above is
unitarizable, which means that for a particular choice of the real form
the representation is unitary. One can show \cite{CLZ03} (Sec.3.2) that for this type of  representations
the generators have the following Hermitian conjugation properties
\beq
\eta E^\dagger_{ab}\eta=E_{ba}
\eeq
where
\beq
\eta={\rm diag}(-1_{M_1},+1_N,+1_{M_2})\;.
\eeq
We see that the representations described above are indeed the representations of ${\rm SU}(M_1,M_2|N)$!
We will examine this property below in Sec.\ref{sl2} for an explicit example.

\subsection{Comparing highest weight with the equation for characters}
We can easily check \eqref{weight242} by extracting the highest weight from
our expression for characters \eq{CharSol}. We want the descendants of the
highest weight to be suppressed which implies $e^{-\alpha_i}\ll 1$. From \eq{alphai}
and \eq{xyident}
we see that this can be
achieved in the limit $|x_3|\gg |x_4|\gg |y_1| \gg |y_2|\gg |y_3|\gg |y_4|\gg  |x_1|\gg |x_2|$.
In this limit, we find from \eq{CharSol}
\beq
T_{1,+s}\simeq\frac{x_1^{s-2} y_1 y_2 y_3 y_4}{x_3 x_4}\;\;,\;\;
T_{1,-s}\simeq\frac{1}{x_3^{s-2} x_4^2}\;\;,\;\;s\ge 2
\eeq
and
\beq
T_{a,+2}\simeq\frac{y_1^a y_2^a y_3^a y_4^a}{x_3^a x_4^a}\;\;,\;\;
T_{a,+1}\simeq\frac{y_1^a y_2^a y_3^a}{x_3^a x_4^a}\;\;,\;\;
T_{a,+0}\simeq\frac{y_1^a y_2^a}{x_3^a x_4^a}\;\;,\;\;
T_{a,-1}\simeq\frac{y_1^a}{x_3^a x_4^a}\;\;,\;\;
T_{a,-2}\simeq\frac{1}{x_3^a x_4^a}
\eeq
which is in  complete agreement with the highest weight \eq{weight242}
after the identification \eq{xyident}.

\subsection{Example: representations of ${\rm GL}(2)$}\la{sl2}

Let us first study this type of representations on the simplest example of \(\frak{gl}(1|0|1)\),
where the simple root is given as $\alpha =\varepsilon_{1}-\varepsilon_{2}$.
The corresponding characters can be calculated from \eqref{KMgen}, similarly to \eqref{oint224}, as follows:

\begin{equation}T_{1,s}\equiv T_s^{(1+1)}=\frac{1}{x_2}\sum_{j={\rm max}(0,-s)}^\infty T^{(1)}_{s+j}(x_1) T^{(1)}_{j}(1/x_2)=
\sum_{j={\rm max}(0,-s)}^\infty x_1^{s+j}x_2^{-j-1}\end{equation}

A simple calculation gives:

\begin{equation}\la{ch11} T_{1,s}\equiv T_s^{(1+1)}(x_1,x_2)=\left\{\bea{cc}
\frac{x_1^{s}}{x_2-x_1}&,\;\;s>0\\
\frac{x_2^{s}}{x_2-x_1}&,\;\;s\le 0
\eea\right.\;.\end{equation}

It is interesting to notice that if we want to satisfy the  Q-system  \eqref{Hirota} with \(T_{1,s}\equiv T_s^{(1+1)}\)we have to   add, after fixing the gauge \(T_{0,s}=1,\,\, -\infty<s<\infty\), another set of representations
\begin{equation}\label{Ta0-rep}T_{a,0}=\frac{x_2^{1-a}}{x_2-x_1}\;.\end{equation}

\begin{figure}[t]
\begin{center}
\includegraphics[scale=0.5]{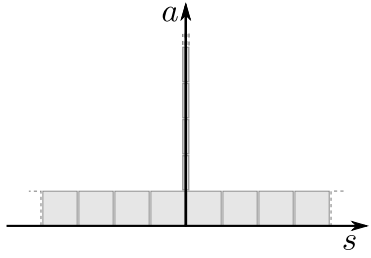}
\end{center}
\caption{A degenerate ``T-hook"  for the ${\gl}(1|0|1)$ representations.
}\label{FatHook2}
\end{figure}

Curiously, although we are not dealing here with a supergroup,  the characters as solutions of the  Q-system live in a  T-hook, though having  zero  width
in the vertical strip (see
Fig.\ref{FatHook2}).

From \eq{weight} with $M_1=M_2=1$ and $N=0$ we have for these representations the highest weight
\begin{align}
\Lambda =
\left\{
\bea{lcc}
-\varepsilon_{1}+|s|\varepsilon_{2}&,\;\;s>0&,\;\;a=1\\
-\varepsilon_{1}-|s|\varepsilon_{1}&,\;\;s<0&,\;\;a=1\\
-a\varepsilon_{1}&,\;\;s=0&,\;\;a>0\\
\eea
\right.
\end{align}
The $s=0,a>0$ case is described by  the same representation as $s<0,a=1$
under the identification $a=1-s$ which is consistent with \eq{Ta0-rep} and \eq{ch11}. In what follows we consider only $a=1$ case.
Note that for  both representations the Dynkin label
is $-|s|-1$. For the unimodular case $x_1=1/x_2$
 the characters \eq{ch11} are indeed equal $T_{1,s}=T_{1,-s}$
because $\varepsilon_{1}+\varepsilon_{2}=0$.
To build the highest weight representation we follow the standard procedure.
We introduce the following combinations of  generators
\beqa
&&h_3\equiv E_{11}-E_{22}\;\;,\;\;h_0\equiv E_{11}+E_{22}\\
&&h_+\equiv E_{12}\;\;,\;\;h_-\equiv -E_{21}
\eeqa
where $h_0$ commutes with all generators and the other commutation relations are
\beq\la{comrel}
\,\,[h_3,h_\pm]=\pm2 h_\pm\;\;,\;\;  [h_-,h_+]=h_3\;.
\eeq
Since $\Lambda(h_0)=s-1$ we have $h_0=(s-1){\rm id}$.
Then $\Lambda( h_3)=-|s|-1$ which implies the following infinite matrix representations of
the generators
\begin{equation}
h_3= \begin{pmatrix}-|s|-1 & 0 & 0 & \cdots \\
0 & -|s|-3 & 0 & \cdots \\
0 & 0 & -|s|-5 &  \cdots \\
 \cdots &  \cdots &  \cdots &  \cdots \\
\end{pmatrix}\;,
\end{equation}
and
\begin{multline}
h_+=\begin{pmatrix}0 & a_0 & 0 & 0 & \cdots \\
0 & 0 & a_1 & 0 & \cdots \\
0 & 0 & 0 & a_2 & \cdots \\
0 & 0 & 0 & 0 & \cdots \\
\cdots & \cdots & \cdots & \cdots & \cdots \\
\end{pmatrix}
, \qquad
h_-=
\begin{pmatrix}0 & 0 & 0 & 0 & \cdots \\
-a_0 & 0 & 0 & 0 & \cdots \\
0 & -a_1 & 0 & 0 & \cdots \\
0 & 0 & -a_2 & 0 & \cdots \\
\cdots & \cdots & \cdots & \cdots & \cdots \\
\end{pmatrix}
,
\\ a_j=i \sqrt{(j+1)(j+|s|+1)}
\end{multline}
One can see that the \(\gl(2)\) commutation relations \eq{comrel} are satisfied.
We notice that we are dialing with the unitary representations of $\frak{u}(1,1)$
since we have\footnote{For $\frak{u}(2)$ one should have  $h^s_+=-(h_-^s)^\dagger$.}
\beq
 h^s_+=(h_-^s)^\dagger\;.
\eeq

Now we can compute the character for these representations.
 Using the identification \eq{xyident} we get
\begin{multline}\la{good}
{\rm tr}\exp\lb\log x_2 E_{11}+\log x_1 E_{22}\rb=
{\rm tr}\exp\lb\frac{h_0}{2}\log x_1x_2+\frac{h_3}{2}\log \frac{x_2}{x_1}\rb=\\
=\sum_{j=1}^\infty\exp\lb\frac{s-1}{2}\log x_1x_2-\frac{|s|+2j-1}{2}\log \frac{x_2}{x_1}\rb
=\frac{x_2^{\frac{s-|s|}{2}}x_1^{\frac{s+|s|}{2}}}{x_2-x_1},
\end{multline}
which  perfectly agrees with \eq{ch11} and  with the character formula in \cite{CLZ03}.
More formally, one can write the character as
\beq
\sum_{n=0}^{\infty} e^{\Lambda - n\alpha}=\frac{e^{\Lambda}}{1-e^{-\alpha}}
\eeq
and then put $x_{1}=e^{\varepsilon_{2}}$ and $x_{2}=e^{\varepsilon_{1}}$ similarly to \eq{xyident} to get \eq{good}.

Now let us consider the case ${\gl(0|2|0)}$ with $M_1=M_2=0$ and $N=2$. From \eq{weight} we have
\begin{align}
\Lambda_s =
\left\{
\bea{ll}
a\varepsilon_{1}+a\varepsilon_{2}&,\;\;s=0\\
a\varepsilon_{1}&,\;\;s=-1\\
0&,\;\;s=-2\\
\eea
\right.
\end{align}
and $\alpha=\varepsilon_1-\varepsilon_2$. The normalized Dynkin label
$2(\Lambda|\alpha)/(\alpha|\alpha)$ is $a$ for $s=-1$
and $0$ for $s=0,-2$. This means that the $s=-1$ case is the $a+1$ dimensional unitary representation of $\frak{u}(2)$.

\subsubsection{ Quantum $GL(1+1)$}

We can easily generalize this character solution of the \(GL(2)\) Hirota equation \eqref{Tsystem} with the T-hook boundary conditions Fig.\ref{FatHook2} to the full spectral parameter dependent solution of \eqref{Hirota}. Let us choose a gauge \( T_{0,s}=1,\quad -\infty <s<\infty,\,T_{1,0}=1\) and denote \( T_{1,s>0}(u)=\frac{\Phi_R^+}{\Phi_R^-},\quad,\,T_{1,s<0}(u)=\frac{\Phi_L^+}{\Phi_L^-}\,,\,\,T_{2,0}=\frac{\Psi^+}{\Psi^-}\). The only non-trivial Hirota equation expresses \(\Psi\) through two independent
functions parameterizing the solution:
\begin{equation}
\xi(u):=\frac{\Psi^+}{\Psi^-}=1-\frac{\Phi_R^+}{\Phi_R^-}\frac{\Phi_L^+}{\Phi_L^-}
\end{equation}
which can be solved as
\begin{equation}
\Psi(u+iK)=\Psi(u)\prod_{k=0}^{K-1}\xi(u+i(k+1/2))
\end{equation}
The rest of the non-trivial T-functions are easily expressed from the Hirota
equation \eqref{Tsystem} which has in these cases only two terms and represents
the discrete Laplace equation:
\begin{eqnarray}
T_{a>0,0}(u)&=& \frac{\Psi(u+i(a-1)/2)}{\Psi(u-i(a-1)/2)},\\
T_{1,s>0}(u)&=&\frac{\Phi_R(u+is/2)}{\Phi_R(u-is/2)},\\
T_{1,s<0}(u)&=& \frac{\Phi_L(u+is/2)}{\Phi_L(u-is/2)}.\\
\end{eqnarray}

Thus we parameterized the general solution in terms of two functions \(\Phi_{L,R}(u)\). It would be interesting to understand whether this system describes any
integrable sigma model, with appropriate analyticity properties for these functions.




\section{Solution of Y-system in the scaling limit  and quasiclassical strings}\la{sec:3old}
In this section we recall the construction   \cite{Gromov:2009tv}
of the general solution of  the  AdS/CFT  Y-system for an arbitrary state,
to the leading wrapping order  \({\cal\ O} (e^{-cL})\).
An important feature of this solution is that it is in one to one
correspondence with the large \(L\) spectrum of the theory
given by asymptotic Bethe ansatz of \cite{Beisert:2006ez}.
It can be used to establish a link between
an exact solution of the Y-system, or Hirota equation, and the corresponding  state of the theory.
Then we use this asymptotic solution of
 the Y-system to identify our \({\rm SU}(2,2|4)\) character solution \eqref{CharSol}
 with the full quasiclassical result of \cite{Gromov:2007ky}  containing the complete one-loop approximation
 of the Metsaev-Tseytlin sigma-model around any classical finite gap solution.
In  \cite{Gromov:2009tq} it was proposed to read off the quantum numbers
of a state by matching the behavior at large $a$ and large $s$
with its asymptotic large \(L\) solution.
In this section we apply this procedure for a general classical state, using our new general character solution of the Y-system in the strong coupling limit.
For each classical finite gap solution we find the corresponding
solution of Y-system.
Then we show that this general classical Y-system solution
has the same structure as the one arising in the direct
one-loop quasiclassical string analysis\footnote{
Though the structures stemming from these two very different approaches, the Y-system on the one hand
and the quasiclassical quantization of the finite gap algebraic curve on the other hand
 (confirmed by the direct one-loop computations in the string  functional integral) will be convincingly identical,
for the complete
comparison one should study  the
exact Bethe equations for auxiliary roots \(u_{a,j}\) (see below). We postpone the detailed analysis for the future work.}.
\subsection{Asymptotic solution in scaling limit}\label{Tscaling}
Let us remind the asymptotic large \(L\) solution  of the Y-system  \cite{Gromov:2009tv}   compatible with the ABA of  \cite{Beisert:2006ez,Beisert:2005fw}.   In the context of ABA the states are parameterized
by $7$ types of Bethe roots (one for each node of the $\frak{psu}(2,2|4)$
Dynkin diagram). We denote them $u_{a,j}$ where the roots of each
type $a=1,\dots,7$ are labeled by the index $j=1,\dots,K_a$.
The auxiliary  roots $(u_{1,j},u_{2,j},u_{3,j})$
and $(u_{5,j},u_{6,j},u_{7,j})$ describe the ``magnons" of the left and right $\frak{su}(2|2)$
subalgebras. The middle node roots $u_{4,j}$ are  momentum carrying and
can be viewed as the rapidities of the inhomogeneities of an
$\frak{su}_{\rm L}(2|2)\oplus \frak{su}_{\rm R}(2|2)$ spin chain.
The corresponding $\frak{su}_{\rm L}(2|2)$
S-matrix, satisfying the Yang-Baxter equation,
was obtained in \cite{Beisert:2005tm}. For a given $K_4$
one can construct the full \(\frak{psu}(2,2|4)\) transfer matrix from this
S-matrix. Its eigenvalues are also parameterized
by the auxiliary roots $(u_{1,j},u_{2,j},u_{3,j})$
for the left wing and $(u_{5,j},u_{6,j},u_{7,j})$
for the right wing. In the simplest fundamental \(\frak{su}_{\rm R}(2|2)\)    representation
for the auxiliary transfer-matrix the  eigenvalues are \cite{Gromov:2009tv}
\beq
{\bf T}^{\rm R}_{1,1}(u)=- {\bf T}^{\rm R,1}_{1,1}(u)+{\bf T}^{\rm R,2}_{1,1}(u)+ {\bf T}^{\rm R,3}_{1,1}(u)- {\bf T}^{\rm R,4}_{1,1}(u)
\eeq
where
\begin{eqnarray*}
{\bf T}^{\rm R,1}_{1,1}(u)&=&
\frac{Q_1^{-}}{Q_1^+}\prod_{j=1}^{K_4}
\frac{1-1/(x^+x_{4,j}^-)}{1-1/(x^+x_{4,j}^+)}\frac{x^--x_{4,j}^-}{x^--x_{4,j}^+},\qquad
{\bf T}^{\rm R,2}_{1,1}(u)=\frac{Q_1^{-} Q_2^{++}}{Q_1^+Q_2  }\prod_{j=1}^{K_4}
\frac{x^--x_{4,j}^-}{x^--x_{4,j}^+},\\
\quad  {\bf T}^{\rm R,3}_{1,1}(u)&=&\frac{Q_2^{--} Q_3^+}{Q_2Q_3^-}\prod_{j=1}^{K_4}
\frac{x^--x_{4,j}^-}{x^--x_{4,j}^+},
\,\,\qquad\qquad\qquad {\bf T}^{\rm R,4}_{1,1}(u)=\frac{Q_3^+ }{Q_3^-  }
\end{eqnarray*}
with the \(Q\)-functions and Zhukowsky \(x(u)\) variables defined as follows
\beqa
\nonumber&&Q_a=Q_a(u)=\prod_{j}^{K_a}(u-u_{a,j})\\
&&x=x(u)=\frac{1}{2}\lb u/g+i\sqrt{4-u^2/g^2}\rb\\
\nonumber&&x_{a,j}=\frac{1}{2}\lb
        u_{a,j}/g+\sqrt{u_{a,j}/g-2}\;\sqrt{u_{a,j}/g+2}\rb\;\;.
\eeqa
 Similar formulas are true for   \(\frak{su}_{\rm L}(2|2) \) transfer matrix \({\bf T}^{\rm L}_{1,-1}\) and we omitted the \(R,L\) superscripts by \(Q\)-functions. For more general representations in auxiliary space
with rectangular $a\times s$ Young diagrams one can use the following
generating function \cite{Tsuboi:1997iq,Beisert:2006qh,Kazakov:2007fy,Gromov:2009tv}
\begin{equation}
\displaystyle{\cal W}_{\rm R}=\!\!
\left[1\!-\!{\bf T}^{\rm R,1}_{1,1} D\right]\!\cdot\!\left[1\!-\!
{\bf T}^{\rm R,2}_{1,1} D\right]^{-1}\!\cdot\!\displaystyle\left[1\!-\! {\bf T}^{\rm R,3}_{1,1} D\right]^{-1}\!\cdot\!
\left[1\!-\!{\bf T}^{\rm R,4}_{1,1} D\right]\,\, , \,\qquad D=e^{-i\partial_u}
\label{eq:gen}
\end{equation}
To generate the corresponding \({\bf T}_{a,1}\) and \({\bf T}_{1,s}\)  one should expand the  above functional in a formal series in $D$ and commute all $D$'s to the right
\beq
{\cal W}_{\rm R}=\sum_{s=0}^\infty
{\bf T}_{1,s}^{{\rm R}}\lb u+i\tfrac{1-s}{2}\rb D^s \,\, , \,\,
{\cal W}_{\rm R}^{-1}=\sum_{a=0}^\infty (-1)^a
{\bf T}_{a,1}^{{\rm R}}\lb u+i\tfrac{1-a}{2}\rb D^a\;.
\eeq
In our gauge ${\bf T}^{\rm R}_{a,0}={\bf T}^{\rm R}_{0,s}=1$ and
the boundary ${\bf T}^{\rm R}_{a,2},\;{\bf T}^{\rm R}_{2,s}$ (corresponding to typical
representations) can
be easily found from the Hirota equation \eqref{Tsystem}.
One can also write similar equations for the left wing,
replacing $(Q_1,Q_2,Q_3)$ by $(Q_7,Q_6,Q_5)$.
In order to match the ABA equations (valid at  large \(L)\)    it was proposed in \cite{Gromov:2009tv}  to relate these eigenvalues of  transfer matrices
to the Y-functions in the following way
\beq\la{YbtoT}
{\bf Y}_{\fp_a}=\frac{{\bf T}^+_{a,1}{\bf T}^-_{a,1}}{{\bf T}_{a+1,1}{\bf T}_{a-1,1}}-1\;\;,\;\;
1/{\bf Y}_{\fb_s}=\frac{{\bf T}^+_{1,s}{\bf T}^-_{1,s}}{{\bf T}_{1,s+1}{\bf T}_{1,s-1}}-1\;,
\eeq
\beqa
&&{\bf Y}_{\fm_a}\simeq {\bf T}^{\rm R}_{a,1}{\bf T}^{\rm L}_{a,1}\prod_{n=-\frac{a-1}{2}}^{\frac{a-1}{2}}
\Phi(u+in)
\eeqa
where
\beqa
\Phi(u)&=&\lb\frac{x^{-}}{x^{+}}\rb^{L}\prod_{j=1}^{K_4}\sigma^2(u,u_{4,j})\frac{x_{4,j}^+}{x_{4,j}^-}\frac{(1/x^+-x_{4,j}^-)(x^--x_{4,j}^+)}{(1/x^--x_{4,j}^+)(x^+-x_{4,j}^-)}
\times\\ \nn&\times&\prod_{j=1}^{K_1}\frac{1/x^+-x_{1,j}}{1/x^--x_{1,j}}
\prod_{j=1}^{K_3}\frac{1/x^--x_{3,j}}{1/x^+-x_{3,j}}
\prod_{j=1}^{K_5}\frac{1/x^--x_{5,j}}{1/x^+-x_{5,j}}
\prod_{j=1}^{K_7}\frac{1/x^+-x_{7,j}}{1/x^--x_{7,j}}\;.
\eeqa
Here \(\s(u,v)\) is the  dressing factor of  \cite{Beisert:2006ez}. This asymptotic solution of the Y-system was constructed
in such a way as to fit the ABA equations of  \cite{Beisert:2006ez,Beisert:2005fw},
which take for the middle nodes the form
\begin{equation}
{\bf Y}^\ph_{\fm_a}(u_{4,j})\simeq -1\;.
\end{equation}

\subsection{Asymptotic T-functions for the entire T-hook}
To do the comparison of our character solution \eqref{CharSol}  to the asymptotic solution of the AdS/CFT Y-system described in the previous section, we have to find the T-functions not only for the
\(\frak{su}_{L,R}(2|2)\) wings but also for the middle nodes. As was mentioned in \cite{Gromov:2009tv} it is possible to pack these two sets of
${\bf T}_{a,s}^{\rm R},{\bf T}_{a,s}^{\rm L}$ with $a,s$ belonging to the
$\frak{su}_{L,R}(2|2)$ fat hooks (Fig.\ref{FatHook}) into one $\frak{psu}(2,2|4)$ T-hook
(Fig.\ref{T-Hook}a).  This, however,
necessarily involves a change of the gauge for T-functions so that at least
one of the wings would be exponentially  suppressed  at large length
$L$, similarly to the example of ${\rm SU}(2)$ principal chiral
field considered in \cite{Gromov:2008gj,Balog:2003yr}. We define
\beqa
{\bf T}_{a,+s}&=&{\bf T}_{a,s}^{\rm R}\prod_{m=-\frac{s-1}{2}}^{\frac{s-1}{2}}\prod_{n=-\frac{a-1}{2}}^{\frac{a-1}{2}}
\Phi^{\rm R}(u+in+im)\;\;,\;\;s>0\nn\\
{\bf T}_{a,+0}&=&1\;\;,\\
{\bf T}_{a,-s}&=&{\bf T}_{a,s}^{\rm L}\prod_{m=-\frac{s-1}{2}}^{\frac{s-1}{2}}\prod_{n=-\frac{a-1}{2}}^{\frac{a-1}{2}}
\Phi^{\rm L}(u+in+im)\;\;,\;\;s>0\nn
\eeqa
where we split the factor $\Phi(u)$ into two
{\small
\begin{align}
\begin{split}
\Phi^{\rm R}(u)\!&=\!\lb\frac{x^{-}}{x^{+}}\rb^{\tfrac{L}{2}}\!\prod_{j=1}^{K_4}\sigma(u,u_{4,j})\sqrt{\frac{x_{4,j}^+}{x_{4,j}^-}
\frac{(\tfrac{1}{x^+}-x_{4,j}^-)(x^--x_{4,j}^+)}{(\tfrac{1}{x^-}-x_{4,j}^+)(x^+-x_{4,j}^-)}}
\prod_{j=1}^{K_1}\frac{\tfrac{1}{x^+}-x_{1,j}}{\tfrac{1}{x^-}-x_{1,j}}
\prod_{j=1}^{K_3}\frac{\frac{1}{x^-}-x_{3,j}}{\tfrac{1}{x^+}-x_{3,j}}\;,\\[5pt]
\Phi^{\rm L}(u)\!&=\!\lb\frac{x^{-}}{x^{+}}\rb^{\tfrac{L}{2}}\!\prod_{j=1}^{K_4}\sigma(u,u_{4,j})\sqrt{\frac{x_{4,j}^+}{x_{4,j}^-}
\frac{(\tfrac{1}{x^+}-x_{4,j}^-)(x^--x_{4,j}^+)}{(\tfrac{1}{x^-}-x_{4,j}^+)(x^+-x_{4,j}^-)}}
\prod_{j=1}^{K_7}\frac{\tfrac{1}{x^+}-x_{7,j}}{\tfrac{1}{x^-}-x_{7,j}}
\prod_{j=1}^{K_5}\frac{\frac{1}{x^-}-x_{5,j}}{\tfrac{1}{x^+}-x_{5,j}}\;,
\end{split}
\end{align}
}so that $\Phi(u)=\Phi^{\rm R}(u)\Phi^{\rm L}(u)$.
Notice that for large $L$ both of these factors are exponentially small.
For $s>0$ and $s<0$ the new ${\bf T}_{a,s}$ are equivalent up to a gauge
transformation to the old ones. This implies that  Hirota
equation is satisfied exactly for $s>0$ and $s<0$.
Moreover, it is easy to check that now Hirota equation is approximately satisfied even for $s=0$, though with an exponential
precision in large $L$ (for a fixed  coupling ). Finally, now we can  write the middle node Y-functions $Y_{\fm_a}$ in terms of these
${\bf T}$'s in a standard way
\beq
{\bf Y}_{\fm_a}=\frac{{\bf T}_{a,+1}{\bf T}_{a,-1}}{{\bf T}_{a+1,0}{\bf T}_{a-1,0}}\;\simeq {\bf T}_{a,+1}{\bf T}_{a,-1}\,\,.
\eeq

\subsection{Classical  limit}\la{sec:cllim}
Now we will take the classical limit in the asymptotic large  \(L\) solution of the AdS/CFT Y-system  described above.
We remind that the ABA equations were constructed
in \cite{AFS,Beisert:2005fw}   to reproduce   the correct
finite gap algebraic curve of \cite{Beisert:2005bm} in the scaling limit $L\sim K_a\sim \sqrt\lambda$.
In this limit
the Bethe roots are densely distributed along some linear stretches in  the complex $u$-plane. These
stretches can be interpreted as branch cuts of some Riemann surface connecting in various
ways $8$ sheets of this surface. We denote the corresponding
$8$-valued function  as $(\lambda_a,\mu_a)$ where $a=1,2,3,4$. They
 are the eigenvalues of the  monodromy matrix of the classical worldsheet theory. Using notations of  \cite{Beisert:2005bm} one can  define
\begin{equation}\label{plambdamu}
 \lh_a\equiv e^{-i\hat p_a},\;\lt_a\equiv e^{-i\tilde p_a}\end{equation}
  where
\beqa
\bea{l}
\hat p_1 = + \frac{L x/(2g)  +\mathcal{Q}_2 x}{x^2-1} + H_1 + \bar H_3 - \bar H_4\\
\tilde p_1 = + {\frac{L x/(2g) - \mathcal{Q}_1}{x^2-1}} + H_1 - H_2 - \bar H_2 + \bar H_3 \\
\tilde p_2 = + {\frac{L x/(2g)  - \mathcal{Q}_1}{x^2-1}} + H_2 - H_3 - \bar H_1 + \bar H_2 \\
\hat p_2 = + {\frac{L x/(2g)  + \mathcal{Q}_2x }{x^2-1}} - H_3 + H_4 - \bar H_1 \\
\hat p_3 = - {\frac{L x/(2g) + \mathcal{Q}_2x }{x^2-1}}  + H_5 - H_4 + \bar H_7 \\
\tilde p_3 = - {\frac{L x/(2g) - \mathcal{Q}_1}{x^2-1}} - H_6+ H_5 + \bar H_7- \bar H_6  \\
\tilde p_4 = - {\frac{L x/(2g)   -\mathcal{Q}_1}{x^2-1}}  - H_7+ H_6 +\bar H_6 - \bar H_5\\
\hat p_4= - {\frac{L x/(2g) - \mathcal{Q}_2 x }{x^2-1}} - H_7 - \bar H_5 + \bar H_4\,\,.
\eea
\label{eq:p}
\eeqa
 Here the Bethe root resolvents $H_a$ are
 \beq
H_a=\sum_{j=1}^{K_a}\frac{x^2}{x^2-1}\frac{1}{x-x_{a,j}}\;\;,\;\;\bar H_a(x)=H_1(1/x)\;.
\eeq
Expanding \eq{eq:gen} in the scaling limit  one gets\footnote{
For this expansion we assume the spectral parameter to be in the upper
half plane. For the other values an analytical continuation should be performed.
See for example
see  \cite{Gromov:2005gp} for more details of this procedure.}
\beq
{\cal W}^{\rm R}=
\frac{(1-d^{\rm R}\lh_1)(1-d^{\rm R}\lh_2)}{
(1-d^{\rm R}\lt_1)(1-d^{\rm R}\lt_2)}\;\;,\;\;
{\cal W}^{\rm L}=
\frac{(1-d^{\rm L}/\lh_4)(1-d^{\rm L}/\lh_3)}{
(1-d^{\rm L}/\lt_4)(1-d^{\rm L}/\lt_3)}
\eeq
where $d^{\rm R,\rm L}$ are  new formal expansion parameters related to the old $D$ in the
following way
\beqa
d^{\rm R}&=&D\exp\left[-i\lb-\frac{L x/(2g)+x{\cal Q}_2}{x^2-1}-H_4+\bar H_1-\bar H_3\rb\right]
\;\;,\\[4pt]
d^{\rm L}&=&D\exp\left[-i\lb-\frac{L x/(2g)+x{\cal Q}_2}{x^2-1}-H_4+\bar H_7-\bar H_5\rb\right]\;.
\eeqa
The transfer-matrix eigenvalues look in new notations in this limit as follows
\begin{align}
\begin{split}
{\bf T}^{\rm R}_{1,s}&=\frac{ {\lt_1}^{s-1} ( {\lt_1}- {\lh_1})
   ( {\lt_1}- {\lh_2})- {\lt_2}^{s-1} ( {\lt_2}- {\lh_1})
   ( {\lt_2}- {\lh_2})}{ {\lt_1}- {\lt_2}}\left(\frac{d^{\rm R}}{D}\right)^{s}\\
{\bf T}^{\rm R}_{a,1}&=(-1)^a\frac{ {\lh_1}^{a-1} ( {\lh_1}- {\lt_1})
   ( {\lh_1}- {\lt_2})- {\lh_2}^{a-1} ( {\lh_2}- {\lt_1})
   ( {\lh_2}- {\lt_2})}{ {\lh_1}- {\lh_2}}\left(\frac{d^{\rm R}}{D}\right)^{a}
\end{split}
\end{align}
and similarly for ${\bf T}^{\rm L}$.
Noticing that
\beqa
\Phi^{\rm R}(u)&\simeq&
\exp\left[-i\left(\frac{xL/(2g)+x{\cal Q}_2}{x^2-1}+H_4-\bar H_1+\bar H_3\right)\right]
=\frac{D}{d^{\rm R}}\;,\\[4pt]
\Phi^{\rm L}(u)&\simeq&
\exp\left[-i\left(\frac{xL/(2g)+x{\cal Q}_2}{x^2-1}+H_4-\bar H_7+\bar H_5\right)\right]
=\frac{D}{d^{\rm L}}
\eeqa
we see that the global  ${\bf T}$'s  defined in the previous section
are  functions of  $\lambda_a$ only!
\beqa\label{ClassT}
\nn{\bf T}_{2,+s}&=&(\lh_1-\lt_1)(\lh_1-\lt_2)(\lh_2-\lt_1)(\lh_2-\lt_2)\lt_1^{s-2}\lt_2^{s-2}\;\;,\;\;s>1\\
\nn{\bf T}_{1,+s}&=&\frac{ {\lt_1}^{s-1} ( {\lt_1}- {\lh_1})
   ( {\lt_1}- {\lh_2})- {\lt_2}^{s-1} ( {\lt_2}- {\lh_1})
   ( {\lt_2}- {\lh_2})}{ {\lt_1}- {\lt_2}}\;\;,\;\;s>0\\
{\bf T}_{+0,s}&=&1\\
\nn{\bf T}_{a,+0}&=&1\;\;,\;\;a>0\\
\nn{\bf T}_{a,+1}&=&(-1)^a\frac{ {\lh_1}^{a-1} ( {\lh_1}- {\lt_1})
   ( {\lh_1}- {\lt_2})- {\lh_2}^{a-1} ( {\lh_2}- {\lt_1})
   ( {\lh_2}- {\lt_2})}{ {\lh_1}- {\lh_2}}\;\;,\;\;a>0\\
\nn{\bf T}_{a,+2}&=&(\lh_1-\lt_1)(\lh_1-\lt_2)(\lh_2-\lt_1)(\lh_2-\lt_2)\lh_1^{a-2}\lh_2^{a-2}\;\;,\;\;a>1
\eeqa
and for ${\bf T}_{a,-s}\;,\;s>0$ one should replace $\lambda_a\to 1/\lambda_{5-a}$ and
$\mu_a\to 1/\mu_{5-a}$
 in ${\bf T}_{a,+s}$. We recognize  in these formulae the \({\rm U}(2|2) \) super-characters generated by the formula \eqref{KMgen} with \(x_1=\lh_1,x_2=\lh_2,y_1=\lt_1,y_2=\lt_2\).
This also implies that all asymptotic ${\bf Y}$-functions
can be written solely in terms of $\lambda_j,\mu_j$ in the scaling limit.
Hirota equation is satisfied exactly for all nodes except the middle ones, for $s=0$.
Notice that  $\lh_1,\lh_2,\mu_1,\mu_2\sim \Delta=e^{-\frac{Lx}{2g(x^2-1)}}$ are exponentially small for large $L/g$
whereas $\lh_3,\lh_4,\mu_3,\mu_4\sim 1/\Delta$ are exponentially large. One can see that at $s=0$
Hirota equation is satisfied only with $\Delta^{2a}$ precision for the \(a\)-th middle node. Thus for large $a$'s
  this solution
should share the same behaviors with the exact solution.
Now we will compare the classical  solution \eqref{ClassT}    with our  ${\rm SU}(2,2|4)$  characters
  \eq{CharSol}.
\subsection{Asymptotic solution as a limit of ${\rm U}(2,2|4)$ super-characters}
Before relating the above asymptotic solution to
 \eq{CharSol}, we first perform the following gauge transformation
\beq
\tilde T_{a,s}=\lb\frac{x_3 x_4}{y_3y_4}\rb^a T_{a,s}.
\eeq
In this gauge, the factor $\frac{y_3y_4}{x_3 x_4}$ in \eq{oint224}
will disappear. Then under  a natural identification
which will be  formally confirmed in the next section
\beq\la{mapxl}
x_{i}=\lt_i\;\;,\;\;y_{i}=\lh_i\;\;,\;\;i=1,\dots,4
\eeq
we get
for $L/g\gg 1$, when  $\{\lt_1,\lt_2|\lh_1,\lh_2\}\sim \Delta\ll 1$ and $\{\lt_3,\lt_4|\lh_3,\lh_4\}\sim 1/\Delta\gg 1$,
\beq
\tilde T_{a,s}\simeq {\bf T}_{a,s}+{\cal O}(\Delta^{a|s|+2})
\eeq
and ${\bf T}_{a,s}\sim \Delta^{a|s|}$.
Notice that the map \eq{mapxl} in particular implies that the
$T_{a,s}$ have a very simple interpretation from the worldsheet
theory point of view. We remind that
$\{\lt_1,\dots,\lt_4|\lh_1,\dots,\lh_4\}$
are the eigenvalues of the classical monodromy matrix
\beq
\Omega={\rm P}\exp\oint {\cal L}\, d\sigma
\eeq
where ${\cal L}$ is the classical Lax connection constructed as a linear
combination of the worldsheet spacial and temporal components
of the classical \({\rm SU}(2,2|4)\) current with the coefficients depending on a
spectral parameter \cite{Bena:2003wd}. Thus $T_{a,s}$ are simply the characters of the
monodromy matrix!\footnote{This kind of relations is similar to the way
the asymptotic solution is constructed as eigenvalues of transfer matrices
of the asymptotic spin-chain. See also \cite{Beisert:2005di}, App.B and \cite{Beisert:2005bm}, eq.(2.40).}
\beq
T_{a,s}={\rm Str}_{a,s}\Omega\;.
\eeq
\(T_{a,s}\), as well as the monodromy matrix, is thus an explicit functional of the elementary fields
of Metsaev-Tseytlin superstring. One can speculate that at the quantum level a similar relation exists.
Namely, we expect
\beq
T_{a,s}=\langle {\rm state}|{\rm Str}_{a,s}\hat\Omega|{\rm state}\rangle\;.
\eeq
Usually  Hirota equation follows for this kind of objects automatically
provided a Yang-Baxter relation is satisfied for the quantum analog of the Lax connection.
Of course the  details  of this identification could be  complicated
\footnote{It would be interesting
to study this kind of relation at weak coupling where
the spectrum is governed by a $\su(2,2|4)$ generalization
of the rational Heisenberg spin chain whose space of state
is infinite dimensional. For that one may
evaluate the universal $R$-matrix for evaluation representations of the
the super Yangian $Y({\frak{gl}}(4|4))$ based on infinite dimensional representations
of $\frak{u}(2,2|4)$ mentioned in Sec.\ref{AppRep} and use super-symmetric extension
\cite{Kulish:2005qc,Tsuboi:2009ud} of the Bazhanov-Lukyanov-Zamolodchikov construction
of T and Q-operators \cite{Bazhanov:1998dq}.
This should lead to the weak coupling limit of $T_{a,s}$ in physical kinematics.
}
\cite{Mikhailov:2007eg}.

\subsection{Fixing the parameters of the general solution}
In this section we derive the
map \eq{mapxl} in a direct  way, similarly to how it was done for the \(\frak{sl}(2)\) sector
in \cite{Gromov:2009tq}.
The map \eq{mapxl} could be  established by comparing the large $a$ and $s$ asymptotics of \(T\)-functions \eqref{ClassT} and the characters \eqref{CharSol}.
First of all we formally treat \eq{CharSol} as a general solution of  Hirota equation
in the T-hook Fig.\ref{T-Hook}a. Indeed it has $7$ independent gauge invariant parameters
(due to \eq{scT} we can always set, say, $x_1=1$) and this is the maximal
number of independent parameters for a solution in the T-hook up to the  gauge transformations.
\footnote{
For example, one can specify $T_{3,s},\;T_{4,s},\;s=-2,-1,0,1,2$ and $T_{0,1}$ and reconstruct all the others $T_{a,s}$
from these $5+5+1$ functions by means of the Hirota equation. It follows from Hirota equation that
$T_{2,s}=(T^+_{3,s}T^+_{3,s}-T_{3,s-1}T_{3,s+1})/T_{4,s}$. Repeating this procedure it is easy to
reconstruct all $T_{a,s}$ for any $s$ and $a\ge|s|$. Then, along with $T_{0,1}$,
one finds in the same way $T_{a,\pm s}$ for any $a$ and $s>a$.
One should extract then $4$ gauge transformations
$T_{a,s}\to g_1 g_2^a g_3^s g_4^{as} T_{a,s}$ to get
$11-4=7$ parameters.}. Together with the gauge transformations we have $11$ parameters to fix.
We can always choose the gauge $T_{0,s}=1$ which fixes $g_1=g_3=1$ leaving
us with   only $9$ parameters.
We will fix these parameters by comparing the large $a$ and $s$ behavior of the exact $T_{a,s}$ with
the asymptotic ${\bf T}_{a,s}$ and get the map
between
$(x_1,\dots,x_4,y_1,\dots,y_4,g_2)$ and $(\lh_1,\dots,\lh_4,\lt_1,\dots,\lt_4)$.
We have for that precisely $9$ equations
\beqa
&&\lim_{a\to+\infty} \frac{1}{a}\log\frac{T^g_{a,s}}{{\bf T}_{a,s}}=0\;\;,\;\;s=-2,-1,0,1,2\\
&&\lim_{s\to+\infty} \frac{1}{s}\log\frac{T^g_{a,s}}{{\bf T}_{a,s}}=0\;\;,\;\;a=1,2\\
&&\lim_{s\to-\infty} \frac{1}{s}\log\frac{T^g_{a,s}}{{\bf T}_{a,s}}=0\;\;,\;\;a=1,2
\eeqa
where
\beq
T^g_{a,s}=g_2^a T_{a,s}.
\eeq
Using the explicit expression \eqref{CharSol} one finds
\beqa\la{eqTa2}
&&\lim_{a\to+\infty} \frac{1}{a}\log\frac{T^g_{a,+2}}{{\bf T}_{a,+2}}=\log \lb \frac{y_1y_2y_3y_4}{x_3x_4}\frac{g_2}{\lh_1\lh_2}\rb
\eeqa
as well as
\beq\label{Ta0}
\lim_{a\to+\infty} \frac{1}{a}\log\frac{T^g_{a,0}}{{\bf T}_{a,0}}=\log \lb \frac{y_3y_4}{x_3x_4}g_2\rb
\eeq
We had to assume here that
\beq
|y_3y_4|>|y_2 y_4|,|y_1 y_4|,|y_2 y_3|,|y_1 y_3|,|y_1y_2|
\eeq
which is the case asymptotically because $y_3,y_4\sim 1/\Delta$, $y_1,y_2\sim\Delta$, where $\Delta$
is exponentially small for large $L$. From \eqref{eqTa2} and \eqref{Ta0} we obtain
\beq
g_2=\frac{x_3x_4}{y_3y_4}\,,\qquad y_1y_2=\l_1\l_2\;.
\eeq
Similarly
\beq\la{eqTa1}
\lim_{a\to+\infty} \frac{1}{a}\log\frac{T^g_{a,+1}}{{\bf T}_{a,+1}}=
\lim_{a\to+\infty} \frac{1}{a}\log \lb \frac{A_1 y_1^a+A_2 y_2^a}{B_1 \lh_1^a+B_2\lh_2^a}\rb=0\\
\eeq
since generically we have $|\lh_1|>|\lh_2|$ or $|\lh_2|>|\lh_1|$
and the expression for $T^g_{a,s}$ is symmetric under exchange $y_1\leftrightarrow y_2$.
We can assume $|y_1|>|y_2|$ then from \eq{eqTa1} and \eq{eqTa2}
we get $y_1=\lh_1,\;y_2=\lh_2$ or $y_1=\lh_2,\;y_2=\lh_1$. Again due to the symmetry we
can always choose
\beq
y_1=\lh_1,\;y_2=\lh_2\;.
\eeq
In the same way, from  the other equations one fixes uniquely
(up to a trivial interchange symmetries) the rest of relations, to obtain finally
\beq
y_i=\lh_i\;\;,\;\;x_i=\lt_i\;\;,\;\;i=1,\dots,4\,\,
\eeq
which is the map we conjectured  in the previous section on the basis of
the asymptotic solution.
Notice that in contrast  to \cite{Gromov:2009tq} we fixed the parameters
of the general solution by comparing the large $a,s$ limit of the T-functions
rather than the Y-functions. It has an  advantage because one does not need to rely
on  TBA equations {\it at all} in this consideration.
From this point of view the  TBA approach is only one of the numerous
tests of the fundamental Hirota dynamics. The asymptotic solution and the T-hook boundary conditions for
Hirota dynamics of \cite{Gromov:2009tv} are the only two fundamental blocks needed to find the exact T-functions
and thus to find the AdS/CFT spectrum.
Moreover, in the framework of the integral TBA equations
each state should be subjected to a case-by-case study
since the ``driving" terms could be very different for different
states and could change when
a singularity of the  integrand crosses the integration contours at some values of
parameters
\cite{S1,S2,S3,Gromov:2009zb,KZ} leaving unaffected the equations
in the functional form\footnote{There are usually several equivalent forms integral equations.
They could have a different structures of  singularities which indicates  their artificial, non-physical nature.
}.
Technically, however, one should not ignore completely the TBA equations
since they give a convenient framework for the study of the spectrum. In the next
section we speculate about the possible extension of the TBA equations for  excited states to the other
sectors, using the known exact solution built above for  strong coupling.
\subsection{Magic products}
 $T_{a,s}$ are  rather complicated functions. Moreover they
contain a gauge ambiguity. It was shown in \cite{Gromov:2009tq}
that some particularly important gauge invariant combinations of them
are  relatively simple functions of $\lh_i,\lt_i$. They are some infinite products
of $Y_{a,s}$  called in \cite{Gromov:2009tq} as  ``magic" products. Here
we give their generalizations from the \({\frak{sl}}(2)\) sector to the full theory.
 All of them can be verified simply using the Hirota equation \eqref{Hirota}.  There
exists a couple of relatively simple ``magic" products
\beqa
e^{{\cal M}_F^+}&=&\frac{1}{Y_{1,+1}Y_{2,+2}}\prod\limits_{a=1}^\infty(1+Y_{a,0})=\frac{\lh_1\lh_2}{\lt_1\lt_2}\;\;,\;\;\\
e^{{\cal M}_F^-}&=&\frac{1}{Y_{1,-1}Y_{2,-2}}\prod\limits_{a=1}^\infty(1+Y_{a,0})=\frac{\lt_3\lt_4}{\lh_3\lh_4}
\eeqa
and a bit more complicated ones
\beqa\la{eqM0}
e^{{\cal M}_0}&\equiv&\prod_{a=1}^{\infty}(1+Y_{a,0})^a
=e^{{\cal N}_*}\;\;,\;\;{\cal N}_*\equiv\sum_{i=1,2}\sum_{j=3,4}\log\frac{(1-\lt_i/\lh_j)(1-\lh_i/\lt_j)}{(1-\lt_i/\lt_j)(1-\lh_i/\lh_j)}\;,
\eeqa
and
\beqa\la{eqM1}
&&e^{-{\cal M}_{+}^+}\equiv e^{-{\cal M}_0}\prod_{a=1}^\infty\lb{\frac{1+Y_{a,+1}}{1+{\bf Y}_{a,+1}}}\rb^a
=e^{-{\cal N}_{\hat 2*}}
\;\;,\;\;|\lh_1|\!<\!|\lh_2|\;,
\\
&&e^{-{\cal M}_{+}^-}\equiv e^{-{\cal M}_0}\prod_{a=1}^\infty\lb{\frac{1+Y_{a,-1}}{1+{\bf Y}_{a,-1}}}\rb^a
=e^{-{\cal N}_{*\hat 3}}
\;\;,\;\;|\lh_3|\!<\!|\lh_4|\,
\eeqa
where
\beqa\la{eqN}
&&{\cal N}_{\hat i*}\equiv\sum_{j=3,4}\log\frac{(1-\lh_i/\lt_j)}{(1-\lh_i/\lh_j)}\;\;,\;\;
{\cal N}_{\tilde i*}\equiv\sum_{j=3,4}\log\frac{(1-\lt_i/\lh_j)}{(1-\lt_i/\lt_j)}\;,
\\ \nn
&&{\cal N}_{*\hat j}\equiv\sum_{i=1,2}\log\frac{(1-\lt_i/\lh_j)}{(1-\lh_i/\lh_j)}\;\;,\;\;
{\cal N}_{*\tilde j}\equiv\sum_{i=1,2}\log\frac{(1-\lh_i/\lt_j)}{(1-\lt_i/\lt_j)}\;.
\eeqa

For $|\lh_1|>|\lh_2|$ one can use the $\lh_1\leftrightarrow\lh_2$ symmetry.\footnote{
It is clear that $e^{{\cal M}_{+}^+}$ is not analytic
for real values of the spectral parameter since it follows from \eqref{eq:p} that  on the real axis (between the branch points)
$|\lh_1|=|\lh_2|$ due to the $x\to 1/x$ symmetry. Similar statement is true for  $e^{{\cal M}_{+}^-}$ which
is symmetric under the exchange of $\lh_3$ and $\lh_4$. The analytic continuation under the real axis is given by
${\cal M}_{-}^\pm$.} And finally we define
\beqa\la{eqM2}
&&e^{-{\cal M}_{-}^+}\equiv\frac{1+1/Y_{2,+2}}{1+1/{\bf Y}_{2,+2}}\prod_{a=2}^\infty
\lb{\frac{1+{\bf Y}_{a,+1}}{1+Y_{a,+1}}}\rb^{a-2}
=e^{-{\cal N}_{\hat 1*}}
\;\;,\;\;|\lh_1|\!<\!|\lh_2|\\
&&e^{-{\cal M}_{-}^-}\equiv\frac{1+1/Y_{2,-2}}{1+1/{\bf Y}_{2,-2}}\prod_{a=2}^\infty
\lb{\frac{1+{\bf Y}_{a,-1}}{1+Y_{a,-1}}}\rb^{a-2}=
e^{-{\cal N}_{*\hat 4}}
\;\;,\;\;|\lh_3|\!<\!|\lh_4|\;.
\eeqa
In the next section we will see that these products appear in the
equations for the spectrum at the one-loop level and also discuss
them in the context of quasiclassical quantization.

\subsection{One-loop energy and quasiclassical quantization}
In this section we show that the leading classical solution of the Y-system
allows to make a rather nontrivial comparison with the direct
worldsheet quasiclassical one-loop  quantization, technically entirely based  on the algebraic curve of the classical finite gap solution of \cite{Beisert:2005bm}.

\subsubsection{Quasiclassical quantization from the algebraic curve}
In this section we briefly remind the idea of one-loop quantization from
the algebraic curve. The algebraic curve allows one to classify the quasi-periodic classical
solutions of the worldsheet sigma model in a transparent  and covariant way.
The algebraic curve is an 8-sheet Riemann surface constructed out of
the eigenvalues $(\lambda_1,\dots,\lambda_4|\mu_1,\dots,\mu_4)$ of the monodromy matrix $\Omega(x)$.
The BMN vacuum corresponds to the ``empty" curve where the eigenvalues have only  singularities at $x=\pm 1$.
In general, there are some branch cuts  connecting the sheets.
The ABA is equivalent at the classical level to the algebraic curve and the cuts
can be thought of as  condensates of the Bethe roots. The precise map between
the configuration of the Bethe roots and the curve is given by \eq{eq:p},
where $\lambda_i$ and $\mu_i$ are related to  $8$ quasi-momenta
$p_{\ii}=\{\hat p_i,\tilde p_i\}$ by \eqref{plambdamu}
 where $\ii$ can take the values $\hat 1,\dots,\hat 4$ or $\tilde 1,\dots,\tilde 4$.

As was proposed in \cite{Frolov:2002av} the one-loop corrections to the
classical energies can
be computed by studying the spectrum of quadratic fluctuations $\delta E^{\ii\jj}_n$
around a given classical state
\beq
\delta E=\frac{1}{2}\sum_{(\ii\jj)}\sum_n(-1)^{F_{\ii\jj}}\delta E^{\ii\jj}_n
\eeq
where $\ii\jj$ labels the polarization and can take the following values for bosonic fluctuations
 ($F_{\ii\jj}=0$)
\beq\label{polB}
(\ii\jj)=(\tilde 1,\tilde 3),(\tilde 2,\tilde 3),(\tilde 1,\tilde 4),(\tilde 2,\tilde 4)
\;\; ,\;\;(\hat 1,\hat 3),(\hat 2,\hat 3),(\hat 1,\hat 4),(\hat 2,\hat 4)\;,
\eeq
and for fermionic fluctuations  ($F_{\ii\jj}=1$)
\beq\label{polF}
(\ii\jj)=(\tilde 1,\hat 3),(\tilde 2,\hat 3),(\tilde 1,\hat 4),(\tilde 2,\hat 4)
\;\;,\;\;(\hat 1,\tilde 3),(\hat 2,\tilde 3),(\hat 1,\tilde 4),(\hat 2,\tilde 4)\;.
\eeq

The quadratic fluctuations have a natural interpretation in terms of  some additional  small  cuts  on the curve in the background of the macroscopic cuts
corresponding to the initial classical finite gap solution. For example the fluctuation $(\tilde 1,\tilde 3)$ correspond to
additional small cuts connecting $\tilde p_1$ and $\tilde p_3$. The values of the spectral parameter $x$
where one can add  small cuts are not arbitrary. At these points the sheets of the Riemann surface
should touch each other which imposes
\beq\label{contBAE}
p_{\ii}(x_n^{\ii\jj})-p_{\jj}(x_n^{\ii\jj})=2\pi n\;.
\eeq
Addition of these extra cuts has two different effects:
firstly, the small cuts by themselves carry an energy,
secondly, the roots belonging to the big cuts are now slightly displaced
 which also affects the energy
\beq\label{deltaE}
\delta E=\frac{1}{2}\sum_{(\ii\jj)}\sum_n(-1)^{F_{\ii\jj}}\omega( x^{(\ii\jj)}_n)
+\int_{\cal C} \omega(x)\delta\rho(x)\mathbb{}\; dx
\;\;,\;\;
\eeq
where
$
\omega(x)\equiv\frac{x^2+1}{x^2-1}
$.
Notice that we can convert the sum in \eq{deltaE} into an integral as follows
\beq
\frac{1}{2}\oint\frac{dx}{2\pi i}
 \omega(x)\partial_x\cal N_{*}\;.
\eeq
Indeed $\partial_x\cal N_{*}$ defined in \eq{eqM0} has simple poles with residues $\pm 1$
when the condition \eq{contBAE} is satisfied and thus leads precisely to the sum in \eq{deltaE}.
Next we deform the contour in the integral above so that it goes around the unit circle $|x|=1$.
The integral around the unit circle in  Zhukovski parameterization
\(2z=x+1/x\) as follows
\beq\label{deltaEfinal} \delta E_{\rm direct}=\int_{-1}^{1}\frac{dz}{2\pi }
 \frac{z}{\sqrt{1-z^2}}\partial_z\cal N_{*}\;.
\eeq
It is easy to see that $\cal N_*$ is exponentially small when $L/\sqrt\lambda\gg 1$
and to match  the ABA one should drop it. The sum in \eq{deltaE} is however
not necessarily exponentially small in this limit.
This is because while deforming the contour one should also
take into account that the quasi-momenta in $\cal N_*$
could have various quadratic branch cuts. These branch cuts
get also caught into the deformed contour
and they give rise to  generically non-vanishing in this limit contributions.
However, these contributions could be
absorbed into a redefinition of  $\delta\rho$ \cite{Gromov:2007ky} so that the
one-loop correction becomes\footnote{For the $\frak{su}(2)$ and $\frak{sl}(2)$ subsectors the modification is trivial $\varrho=\rho$.}
\beq
\delta E=\delta E_{\rm direct}
+\int_{\cal C} \omega(x)\delta\varrho(x)\mathbb{}\; dx
\;\;.\;\;
\eeq
The modified density
$\varrho(x)$ is precisely the density of the
momentum carrying roots $u_{4,j}$ in the asymptotic limit $L/\sqrt\lambda\gg 1$ \cite{Gromov:2007ky}.
In our case $L/\sqrt\lambda\sim 1$ but  we can show that  $\varrho$
is also a density of the roots $u_{4,j}$ but for  corrected ABA equations
(see Appendix \ref{AppQC} for more details)
\begin{align}
\begin{split}
\la{eqMidCor}
1&=-\lb\frac{x_{4,k}^{-}}{x_{4,k}^{+}}\rb^{L}\prod_{j=1}^{K_4}\sigma_{}^2(u_{4,k},u_{4,j})\frac{x_{4,j}^+}{x_{4,j}^-}\frac{(1/x_{4,k}^+-x_{4,j}^-)(x_{4,k}^--x_{4,j}^+)}{(1/x_{4,k}^--x_{4,j}^+)(x_{4,k}^+-x_{4,j}^-)}
\times
\\ & \times  \prod_{j=1}^{K_1}\frac{1/x_{4,k}^+-x_{1,j}}{1/x_{4,k}^--x_{1,j}}
\prod_{j=1}^{K_3}\frac{x_{4,k}^+-x_{3,j}}{x_{4,k}^--x_{3,j}}
\prod_{j=1}^{K_5}\frac{x_{4,k}^+-x_{5,j}}{x_{4,k}^--x_{5,j}}
\prod_{j=1}^{K_7}\frac{1/x_{4,k}^+-x_{7,j}}{1/x_{4,k}^--x_{7,j}} \\
&\times \exp\left[-\int_{-1}^{+1}\lb r(x_{4,k},z)({\cal N}_{\hat 2*}+{\cal N}_{*\hat 3})-
r(1/x_{4,k},z)({\cal N}_{\hat 1 *}+{\cal N}_{\hat *4})+2u(x_{4,k},z){\cal N}_{*}\rb dz\right]
\end{split}
\end{align}
where we  use the following kernels
\begin{align}
\nn & r(y,z)=\frac{y^2}{y^2-1}\frac{\partial_z}{2\pi g}\frac{1}{y-x(z)}\;\;,\;\;
u(y,z)=\frac{y}{y^2-1}\frac{\partial_z}{2\pi g}\frac{1}{x^2(z)-1}
\qquad
\\
 & q(y,z)= \frac{1}{y^2-1}\frac{\partial_z}{2\pi g}\frac{x(z)}{x^2(z)-1}\;. \nn
\end{align}
The last factor in \eq{eqMidCor} is exponentially small in this scaling limit  and contains the information about all
wrappings for $L\sim\sqrt\lambda$.
Equations for the auxiliary roots should be modified as well.
These modified equations are presented in the Appendix \ref{AppQC}.

Now we will see how  naturally these structures appear in the expression for the energy of a state
and for the exact finite volume Bethe equations.

\subsubsection{One-loop energy from Y-system}
In this subsection we will show how the energy can be computed from the Y-system with the one-loop accuracy.
Expanding \eq{eq:Energy} at strong coupling we get
\begin{equation}
E=\sum_{j=1}^{K_4}\frac{x_{4,j}^2+1}{x_{4,j}^2-1}-\sum_{a=1}^\infty\int_{-\infty}^{\infty}\frac{dz}{2\pi }
\partial_z \frac{z}{\sqrt{1-z^2}}
\log\left(1+Y_{a,0}\right)^a
+{\cal O}\lb 1/\sqrt\lambda\rb
, \label{eq:Energy0}
\end{equation}
where we introduce a rescaled spectral parameter $z=u/(2g)$.
In the scaling limit the number of Bethe roots $K_4$ should scale as
$\sqrt\lambda$. We see that the first term is of the order $\sqrt\lambda$
and thus contains the classical part of the energy which scales as $\sqrt\lambda$
and also some part of the one loop correction which scales as $\lambda^0$.
The second term is already of the order $\lambda^0$ and contributes only
to the one-loop correction. To evaluate the second term one only needs to know
the leading order $Y_{a,0}$ found in the previous sections.
 As it can be seen from the asymptotic expressions for $Y_{a,0}$
in the considered limit, they are supported on the interval $(-1,1)$.
We assume that the exact $Y_{a,0}$ are suppressed outside this interval
and one can integrate only over this interval in \eq{eq:Energy0}.
Using \eq{eqM0} we immediately recognize that the integral term is
precisely $\delta E_{\rm direct}$ of \eqref{deltaEfinal}  from the algebraic curve computation (and hence related to ABA).

Now we have to define the positions of roots $x_{4,j}$ with a precision \({\cal O}\lb\lambda^0\rb\).
They can be expanded in $1/\sqrt\lambda$
\beq
x_{4,j}=x^0_{4,j}+x^1_{4,j}/\sqrt\lambda+\dots
\eeq
The asymptotic values $x^0_{4,j}$ lead to the classical energy from the first term in \eq{eq:Energy0}
whereas the corrections $x^1_{4,j}$  contribute at one-loop.
Below we show that the
equation for $x_{4,j}$ coincides with  the Bethe equation \eq{eqMidCor}  corrected by  virtual fluctuations.
For that one can speculate about a possible form of
TBA equations for  excited states outside the $\frak{sl}(2)$ sector\footnote{We would like to thank P.Vieira
for the  discussion on some of the points raised in this section.}
and extend the consideratiosn of \cite{Gromov:2009tq} to the full theory.

For the $\frak{sl}(2)$ sector the TBA equation for the $Y_{1,0}$ was proposed in \cite{Gromov:2009bc}
to be of the form
\beq
\log Y_{1,0}=\sum_{m=1}^\infty{\cal T}_{1,m}*\log(1+Y_{m,0})
+\sum_{m=1}^\infty 2{\cal R}^{(10)}\cut K_{m-1}*\log(1+Y_{m,1})+i\Phi
\eeq
where $*$ and $\cut$ denote some convolutions, ${\cal T}_{1,m},\;{\cal R}^{(10)}$
are the kernels
  and $\Phi$ is a potential term, all  defined in \cite{Gromov:2009bc}.
This equation is especially important for us since it allows to find the corrected
Bethe equation for the middle node by doing  analytic continuation to the physical
sheet, like in \cite{Gromov:2009zb}.
For the $\frak{sl}(2)$ sector one had $Y_{a,+s}=Y_{a,-s}$ and thus a natural generalization
is
\beqa
\log Y_{1,0}=\sum_{m=1}^\infty{\cal T}_{1,m}*\log(1+Y_{m,0})
+\sum_{m=1}^\infty {\cal R}^{(10)}\cut K_{m-1}*\log(1+Y_{m,+1})(1+Y_{m,-1})+i\tilde\Phi\nn\;.
\eeqa

In \cite{Gromov:2009bc} the potential terms were inferred by recovering  the ABA
type contributions from the kernels of  TBA equations for the vacuum \cite{Bombardelli:2009ns,Gromov:2009bc,Arutyunov:2009ur}. As was shown in \cite{Gromov:2009tq},
technically it is very convenient (and should be also useful for numerics) to subtract the equation
satisfied by the asymptotic solution from the above exact equation, to cancel the potential terms
\footnote{Strictly speaking the driving terms in the exact and asymptotic solution
could be slightly different. We assume that this difference is suppressed.}
\beq
\log \frac{Y_{1,0}}{{\bf Y}_{1,0}}=\sum_{m=1}^{\infty}{\cal T}_{1,m}*\log(1+Y_{m,0})
+\sum_{m=1}^\infty {\cal R}^{(10)}\cut K_{m-1}*\log\lb\frac{1+Y_{m,+1}}{1+{\bf Y}_{m,+1}}\frac{1+Y_{m,-1}}{1+{\bf Y}_{m,-1}}\rb\;.
\eeq
Another advantage of this trick at strong coupling is that
the $Y$'s outside the interval $[-1,1]$ in the rescaled variable $z=u/(2g)$
where $g=\frac{\sqrt\lambda}{4\pi}$ should coincide with the
asymptotic ${\bf Y}$'s since the middle nodes $Y_{a,0}$ are exponentially small for these values
of the spectral parameter in the mirror kinematics. Thus in all convolutions
we can restrict the integrations to the interval $[-1,1]$ where our solution of Q-system is valid.
Making the analytic continuation to the physical sheet we get \cite{Gromov:2009zb}
\beqa
\nn\log \frac{Y^\ph_{1,0}}{{\bf Y}^\ph_{1,0}}&=&\sum_{m=1}^\infty{\cal T}^{\ph,\mir}_{1,m}*\log(1+Y_{m,0})
+\sum_{m=1}^\infty K^-_{m-1}*\log\lb\frac{1+Y_{m,+1}}{1+{\bf Y}_{m,+1}}\frac{1+Y_{m,-1}}{1+{\bf Y}_{m,-1}}\rb\\
&+&\sum_{m=2}^\infty \lb{\cal R}^{(10)\ph,\mir}-{\cal B}^{(10)\ph,\mir}\rb* K_{m-1}*\log\lb\frac{1+Y_{m,+1}}{1+{\bf Y}_{m,+1}}\frac{1+Y_{m,-1}}{1+{\bf Y}_{m,-1}}\rb\\
&+&{\cal R}^{(10)\ph,\mir}*\log\lb\frac{1+Y_{1,+1}}{1+{\bf Y}_{1,+1}}\frac{1+Y_{1,-1}}{1+{\bf Y}_{1,-1}}\rb \nn\\ \nn
&-&{\cal B}^{(10)\ph,\mir}*\log\lb\frac{1+1/Y_{2,+2}}{1+1/{\bf Y}_{2,+2}}\frac{1+1/Y_{2,-2}}{1+1/{\bf Y}_{2,-2}}\rb\;\nn
\eeqa
where the last two terms appeared from converting the convolution around the
B-cycle $\cut$ into two usual integrals $*$ over $[-1,1]$.
Now we simply use the strong coupling expansion of the kernels at large $g$
from \cite{Gromov:2009tq}\footnote{Here we use the equivalent form of
TBA equations for the massive nodes which differs by zero total momentum from the
ones in \cite{Gromov:2009tq}. This requires simultaneous redefinition of the kernel ${\cal T}^{\ph,\mir}_{1,m}(z_k,w)$
and the free terms like in \cite{Gromov:2009at}. As a result the strong coupling expansion of ${\cal T}^{\ph,\mir}_{1,m}(z_k,w)$
is a bit different from \cite{Gromov:2009at}.}
\beqa
\nn{\cal R}^{(10)\ph,\mir}(z_k,w)&\simeq& r(x_k,w)\;,\\
\nn{\cal B}^{(10)\ph,\mir}(z_k,w)&\simeq& r(1/x_k,w)\;,\\
{\cal T}^{\ph,\mir}_{1,m}(z_k,w)&\simeq&-m\[2r(x_k,w)+2u(x_k,w)\]\;,\\
\nn K_{m}(z_k-w)&\simeq&
\delta(w-z_k)+m \[r(x_k,w)+ r(1/x_k,w)\]\;,
\eeqa
which leads to
\beqa
\nn\log \frac{Y^\ph_{1,0}}{{\bf Y}^\ph_{1,0}}
&=&r(x_k,w)\log\prod_{m=1}^\infty\lb\frac{1}{(1+Y_{m,0})^{2}}\frac{1+Y_{m,+1}}{1+{\bf Y}_{m,+1}}\frac{1+Y_{m,-1}}{1+{\bf Y}_{m,-1}}
\rb^m\\
&+&r(1/x_k,w)
\log
\frac{1+1/{\bf Y}_{2,+2}}{1+1/Y_{2,+2}}
\frac{1+1/{\bf Y}_{2,-2}}{1+1/Y_{2,-2}}\prod_{m=1}^\infty\lb\frac{1+Y_{m,+1}}{1+{\bf Y}_{m,+1}}\frac{1+Y_{m,-1}}{1+{\bf Y}_{m,-1}}
\rb^{m-2}\\
&-&2u(x_k,w)\log(1+Y_{m,0})^{m}\nn\;.
\eeqa
Using the relations  (\ref{eqM0},\ref{eqM1},\ref{eqM2})  for this kind of products of Y-functions and assuming that the contours of integration are displaced so that $|\lambda_1|<|\lambda_2|$ and
 $|\lambda_3|<|\lambda_4|$ we get
\beqa
\nn\log \frac{Y^\ph_{1,0}}{{\bf Y}^\ph_{1,0}}
&\simeq&\int_{-1}^1\lb-r(x_k,z)\left({\cal N}_{\hat 2 *}+{\cal N}_{*\hat 3}\right)
+r(1/x_k,z)
\left({\cal N}_{\hat 1 *}+{\cal N}_{*\hat 4}\right)-2u(x_k,z){\cal N}_{*}\nn
\rb dz\;.
\eeqa
Since the exact Bethe equation for the momentum carrying node is $Y_{1,0}^{\ph}(u_{4,k})=-1$ we get
the modified asymptotic Bethe equation \eq{eqMidCor} from the previous section.

We see that precisely the same equations for the spectrum appear in both the  Y-system
and  the quasiclassical quantization of algebraic curve. This is a striking confirmation of the correctness of the AdS/CFT  Y-system \cite{Gromov:2009tv}.
To complete the prove that both approaches lead to the same result
one should also study the auxiliary Bethe equations. Whereas it is easy to obtain the modified form of the auxiliary Bethe equations  from the quasiclassical
quantization, it is usually more complicated to see them directly from the Y-system (see e.g. \cite{Gromov:2008gj}). It should
 follow from some analyticity conditions, which can be read off from the asymptotic
solution of the Y-system. Indeed the asymptotic solution is constructed
in terms of the transfer-matrices and the auxiliary Bethe equations are simply the conditions
of  pole cancellations. We postpone the detailed analysis of the auxiliary equations
for the future work.

\section{Conclusions}\la{sec:4old}

One of the main motivations for this work was to find a good test for the Y-system conjectured in
\cite{Gromov:2009tv}, in the situation when the asymptotic Bethe ansatz is essentially inaccurate
due to the presence of all
multiple windings,  but the Y-system is still treatable analytically. The strong coupling limit considered here, in
the situation when \(L\sim \sqrt{\l}\to\infty\), gives such an opportunity.

On the one hand, the all-wrapping corrections originated from the discreetness
of the sum over the fluctuation frequencies \cite{SchaferNameki:2006gk} are present already in the one-loop energy; on the other hand,
the dependence on the spectral parameter \(u\) becomes slow\footnote{this approximation
is not valid in the vicinity of the real axis for $|u|>2g$ but should give the right result on the rest of the complex plane.} in this scaling limit and hence the finite
difference operator could be
neglected in Y-system and in the related Hirota equation. In this case, the Hirota equation simplifies to the form \eqref{Hirota}
 and becomes
similar to the one solved by the  (super)-characters of irreducible representations with
\((a\times s)\) rectangular Young diagrams. This simplified Hirota equation is often called as ``Q-system" in the mathematical literature.

Another motivation of this work was to understand the nature of the \({\rm PSU}(2,2|4)\) representations  which enter through the \(a,s\) variables into the Q-system. The full superconformal group was usually difficult to identify  because of the light cone gauge in which this superstring theory is studied.  On this way, we managed to find a general solution of such a Y-system with no shifts with respect to the spectral
parameter, with the  AdS/CFT type boundary conditions: Y-functions are defined within the  T-hook
Fig.\ref{T-Hook}b  in the representation space of
\((a\times s)\) rectangular Young diagrams.  The solution we found   appears to define
super-characters of certain unitary infinite-dimensional representations of \({\rm SU}(2,2|4)\).
Comparing the  \(a,s\to \infty\) asymptotics of these super-characters with those of the
asymptotic solution we uniquely identified the
parameters of these super-characters  with the eigenvalues of the
classical monodromy matrix or equivalently with 8 quasi-momenta of an arbitrary finite gap solution.
Importantly, to build the classical solution of the Y-system we use only the data from \cite{Gromov:2009tv}
with no input from the TBA approach.
This reduces drastically the number of assumptions we have to adopt.

Then we show that using this leading order strong coupling solution
of the Y-system we reproduce the equations arising in the worldsheet quasiclassical
quantization procedure. We observe that  precisely the same
structures involving different combinations of quasi-momenta follow from the Y-system
in a very nontrivial way.
At this stage we assume some natural generalization of the TBA equations
for the excited states originally proposed for the $\frak{sl}(2)$ subsector \cite{Gromov:2009bc}.

Probably one of the most interesting problems left is  the derivation of the explicit ``quantum"
 generalization of our solution \eqref{CharSol}, now for the full \(u\)-dependent T-system
 \eqref{Tsystem}, in terms of certain Wronskian-type determinant expressions.
 Unlike the   solution given in \cite{Hegedus:2009ky} using the B\"acklund techniques of \cite{Kazakov:2007fy}, the one we mean here would not contain,
 similarly to  \eqref{CharSol}, any infinite sums, in analogy to the solution of
 \cite{Tsuboi:2009ud} for the super-spin chains.
Then one should fix the parameters of this solution for each given state of the theory.
As was demonstrated in this paper, such  a solution  should be more convenient
for fixing  explicitly the large $a$ and $s$ asymptotics.
 This would be an important step in construction of a finite system of non-linear integral
 equations of a Destri-DeVega type (in analogy with \cite{Gromov:2008gj}) for the AdS/CFT spectrum problem, also exact for any operator and at any coupling \(\l\).
 Apart from its obvious advantages for a numeric analysis, such a finite set of equations,
 extending the observations of this paper to the quantum level, would
 allow us to better understand the  full \({\rm PSU}(2,2|4)\) integrability structure of both sides of the AdS/CFT correspondence   which is somewhat hidden due to the original light cone gauge and the
 related \({\rm SU}(2|2)\times {\rm SU}(2|2)\) setup.

\section*{Acknowledgments}
The work of NG was partly supported by the German Science Foundation (DFG) under
the Collaborative Research Center (SFB) 676 and RFFI project grant 06-02-16786.
The work  of VK was partly supported by  the ANR grants INT-AdS/CFT (BLAN-06-0124-03)  and   GranMA (BLAN-08-1-313695) and the grant RFFI 08-02-00287.
The work  of ZT is supported by
Grant-in-Aid for Young Scientists, B \#19740244 from
The Ministry of Education, Culture, Sports, Science and Technology in Japan.
We  thank N.Beisert, C.Candu, M.R.Douglas, S.Leurent, F.Levkovich-Maslyuk, L.Mazzucato, S.Schafer-Nameki,
M.Staudacher, D.Serban, A.Tseytlin, P.Vieira, D.Volin and K.Zarembo for discussions.
NG and VK thank the
Simons Center for Geometry and Physics,
where a part of the work was done, for the kind hospitality.
ZT thanks Ecole Normale Superieure, LPT,
where a  part of this work was done,  for the hospitality.

\appendix

\section{Mathematica expressions for general solution of Q-system}
\la{AppMath}
{\footnotesize
\verb"S[i_]=((y[i]-x[3])(y[i]-x[4]))/((y[i]-x[1])(y[i]-x[2]));"\\
\verb"Z[i_]=((x[i]-y[1])(x[i]-y[2])(x[i]-y[3])(x[i]-y[4]))/((x[i]-x[3])(x[i]-x[4]));"\\
\verb"t[j_,s_]=Boole[j>s];"\\
\verb"M4[a_,s_]=Table[S[i]^t[j,s+2]y[i]^(j-4-(a+2)t[j,s+2]),{i,4},{j,4}];"\\
\verb"M2[a_,s_]=Table[Z[i]^(1-t[j,a])x[i]^(2-j+(s-2)(1-t[j,a])),{i,2},{j,2}];"\\
\verb"M2n[a_,s_]=M2[a,-s]/.{x[i_]->1/x[5-i],y[i_]->1/y[i]};"\\
\verb"T[a_,s_]:=0/;(a>2&&Abs[s]>2)||a<0;"\\
\verb"T[a_,s_]:=(((-1)^(a s+s)((x[3]x[4])/(y[1]y[2]y[3]y[4]))^(s-a)"\\
\verb"          Det[M4[a,s]])/Det[M4[0,0]])/;a>=Abs[s];"\\
\verb"T[a_,s_]:=Det[M2[a,s]]/Det[M2[0,s]]/;s>=a;"\\
\verb"T[a_,s_]:=((y[1]y[2]y[3]y[4])/(x[1]x[2]x[3]x[4]))^a Det[M2n[a,s]]/Det[M2n[0,s]]/;s<=-a;"\\
}\\
One can see  indeed that  Hirota equation is satisfied (the code is quite time consuming)\\
{\footnotesize
\verb"Hir[a_?NumericQ,s_?NumericQ]:=T[a,s]^2-T[a+1,s]T[a-1,s]-T[a,s+1]T[a,s-1]"\\
\verb"Table[Hir[a,s]//Factor,{a,0,3},{s,-4,4}]"
}\\
{}\\
For the completeness we add also the asymptotic solution from Sec.\ref{sec:cllim}
{\footnotesize\\
\verb"Tb[2,s_]:=(lt[1]-lh[1])(lt[2]-lh[1])(lt[1]-lh[2])(lt[2]-lh[2]) lt[1]^(s-2) lt[2]^(s-2)/;s>1"\\
\verb"Tb[1,s_]:=((lt[1]^(s-1)(lt[1]-lh[1])(lt[1]-lh[2])-lt[2]^(s-1)(lt[2]-lh[1])(lt[2]-lh[2]))"\\
\verb"          /(lt[1]-lt[2]))/;s>0"\\
\verb"Tb[a_,2]:=(lh[1]-lt[1])(lh[1]-lt[2])(lt[1]-lh[2])(lt[2]-lh[2]) lh[1]^(a-2) lh[2]^(a-2)/;a>1"\\
\verb"Tb[a_,1]:=((-1)^a (lh[1]^(a-1)(lh[1]-lt[1])(lh[1]-lt[2])-lh[2]^(a-1)(lh[2]-lt[1])(lh[2]-lt[2]))"\\
\verb"          /(lh[1]-lh[2]))/;a>0"\\
\verb"Tb[0,s_] =1;   Tb[a_,0]:=1/;a >= 0;"\\
\verb"Tb[a_,s_]:=0/;(Abs[s]>2&&a>2)||a < 0;"\\
\verb"Tb[a_,s_]:=(Tb[a,-s]/.{lh[c_]->1/lh[5-c],lt[c_]->1/lt[5-c]})/;s < 0"
}

\section{Derivation of the modified Bethe equations}\la{AppQC}
In this section we show how the effect from  virtual fluctuations,
corresponding to the one-loop quantum corrections,
can be absorbed into a certain modification of the asymptotic Bethe equations.
The algebraic curve can be described by a set of integral equations
for the ``densities", or discontinuities of the quasi-momenta $p_{\bf i}=\{\hat p_{i},\tilde p_{i}\}$ in the following way
\beqa\la{BAEcurve}
\sp_{\bf i}-\sp_{\bf j}=2\pi k\;\;,\;\;x\in {\cal C}^k_{\bf ij}
\eeqa
where $\sp_{\bf i}\equiv (p_{\bf i}(x+i0)+p_{\bf i}(x-i0))/2$.
The fluctuations modify the curve by extra poles at
\beq
x=x_n^{\bf i\bf j}\;\;:\;\;\sp_{\bf i}(x_n^{\bf i\bf j})-\sp_{\bf j}(x_n^{\bf i\bf j})=2\pi n
\eeq
with the residues
\beq
\alpha(x)=\frac{4\pi}{\sqrt\lambda}\frac{x^2}{x^2-1}\;.
\eeq
There are some additional constraints on the asymptotics of the resulting quasi-momenta
 due to the $x\to 1/x$ symmetry. As  is explained in  great detail in
the  paper \cite{Gromov:2007cd} the equations \eq{BAEcurve} are modified by
fluctuations $x_n^{{\bf i},{\bf j}}$
induced by  extra potentials in the following way
\beqa\la{BAEcurveA}
\sp_{\bf l}-\sp_{\bf p}+V^{\ii\jj,n}_{\bf l}-V^{\ii\jj,n}_{\bf p}=2\pi k\;\;,\;\;x\in {\cal C}^k_{\bf lp}
\eeqa
where indexes $\ii,\jj$ characterize the polarization of the fluctuation,
$n$ is roughly a Fourier mode of the fluctuation. These potentials are given by
\beq
\lb
\bea{c}
V_{\hat 1}^{\ii\jj,n}\\
V_{\hat 2}^{\ii\jj,n}\\
V_{\hat 3}^{\ii\jj,n}\\
V_{\hat 4}^{\ii\jj,n}
\eea\rb =\lb
\bea{c}
+1\\
+1\\
-1\\
-1
\eea\rb\frac{x}{x^2-1}\frac{\alpha(x_n^{\ii\jj})}{(x_n^{\ii\jj})^2}
+\lb
\bea{c}
+\delta_{\hat1\ii}\\
+\delta_{\hat2\ii}\\
-\delta_{\hat3\jj}\\
-\delta_{\hat4\jj}\\
\eea
\rb\frac{\alpha(x)}{x-x_n^{\ii\jj}}
-
\lb
\bea{c}
+\delta_{\hat2\ii}\\
+\delta_{\hat1\ii}\\
-\delta_{\hat4\jj}\\
-\delta_{\hat3\jj}\\
\eea
\rb\frac{\alpha(1/x)}{1/x-x_n^{\ii\jj}}\la{Vh} \,,
\eeq
and
\beq\la{Vt}
\lb
\bea{c}
V^{\ii\jj,n}_{\tilde 1}\\
V^{\ii\jj,n}_{\tilde 2}\\
V^{\ii\jj,n}_{\tilde 3}\\
V^{\ii\jj,n}_{\tilde 4}
\eea\rb =
\lb
\bea{c}
-1\\
-1\\
+1\\
+1
\eea\rb\frac{1}{x^2-1}\frac{\alpha(x_n^{\ii\jj})}{x^{\ii\jj}_n}-
\lb
\bea{c}
+\delta_{\tilde1\ii}\\
+\delta_{\tilde2\ii}\\
-\delta_{\tilde3\jj}\\
-\delta_{\tilde4\jj}\\
\eea
\rb\frac{\alpha(x)}{x-x_n^{\ii\jj}}
+
\lb
\bea{c}
+\delta_{\tilde2\ii}\\
+\delta_{\tilde1\ii}\\
-\delta_{\tilde4\jj}\\
-\delta_{\tilde3\jj}\\
\eea
\rb\frac{\alpha(1/x)}{1/x-x_n^{\ii\jj}} \,.
\eeq

For the one-loop shift one should  introduce the fluctuations
with all possible polarizations listed in \eq{polB} and \eq{polF}
and sum over all Fourier modes with a factor $1/2$. As in \eq{deltaE}
we rewrite the sum  as a contour integral around  all fluctuations
$x_n^{\ii\jj}\;\;,\;\;n=-\infty,\dots,\infty$
\begin{align}
\sp_{\bf l}-\sp_{\bf p}+\sum_{\ii\jj}\frac{1}{2}
\oint\frac{dy}{2\pi i}\partial_y\log\sin\frac{p_{\ii}(y)-p_{\jj}(y)}{2}
\lb V^{\ii\jj}_{\bf l}(x,y)-V^{\ii\jj}_{\bf p}(x,y)\rb
=2\pi k\;\;,\quad x\in {\cal C}^k_{\bf lp}
\end{align}
where $V_{\bf p}^{\ii\jj}(x,y)$ is $V_{\bf p}^{\ii\jj,n}$ with $x_n^{\ii\jj}$ replaced by $y$.
Next we deform the integration contour to pass around the unit circle $|y|=1$.
Since the quasi-momenta in general have branch points and the potentials
have poles at $y=x$ we
also  have to add the contributions from all these singularities.
Some of these contributions can be absorbed into
redefinitions of the quasi-momenta $p_\ii\to q_\ii$ so that
$q_\ii$ has the same analytic properties as $p_\ii$.
The  contributions which cannot be absorbed into the redefinition
of $p_\ii$ are called  ``Anomaly" (by  historical reasons, see  \cite{Daul:1993xz,Beisert:2005di,Beisert:2005mq,FS1}).
Now let us  use
\beq
\partial_y\log\sin\frac{p_{\ii}-p_{\jj}}{2}=
\frac{p_{\ii}'-p_{\jj}'}{2}+\partial_y\log\lb 1-e^{-ip_{\ii}+ip_{\jj}}\rb\;.
\eeq
As was shown in \cite{Gromov:2007cd} the first term reflects the contribution of  Hernandez-Lopez (HL) \cite{HL}
phase in ABA
and we get\beqa\la{BAEcurve2}
\sq_{\bf l}-\sq_{\bf p}+{\rm Anomaly}+{\rm HL}-2\pi k=\sum_{\ii\jj}
\oint_{\mathbb U^+}\frac{dy}{2\pi i}\partial_y\log\lb 1-e^{-ip_{\ii}+ip_{\jj}}\rb
\lb V^{\ii\jj}_{\bf l}-V^{\ii\jj}_{\bf p}\rb\;,
\eeqa
where the integration contour ${\mathbb U}^+$ goes along the upper part of the unit circle $|y|=1$.
The last equation is an integral equation for the ``densities" $\varrho=\frac{q(x-i0)-q(x+i0)}{2\pi i}$.

At the same time, it was shown
 in \cite{FS1,Gromov:2007ky} that the ABA  can be written with a one-loop accuracy
 as the following equation for the density of the momentum-carrying
 roots $\varrho$
\beq
\sq_{\bf l}-\sq_{\bf p}+{\rm Anomaly}+{\rm HL}-2\pi k=0\;.
\eeq
Now we clearly see that the discrepancy between the exact one-loop energies
and the prediction of the ABA is due to the integral in the r.h.s.  of \eq{BAEcurve2}.
This last term is responsible to  all-wrapping contributions. We can easily modify the Bethe equations so that the two results agree again at the one-loop level.
For example the equation for the middle node $u_{4,k}$
in the scaling limit becomes
\beq
{\rm ABA}_{u_4}=\exp\left[i\left(\sq_{\tilde 2}-\sq_{\tilde 3}+{\rm Anomaly}+{\rm HL}\right)\right]
\eeq
and to correct it we simply add an extra phase to the r.h.s.
\beq
{\rm ABA}_{u_4}=\exp\left[\sum_{\ii\jj}
\oint_{\mathbb U^+}\frac{dy}{2\pi}\partial_y\log\lb 1-e^{-ip_{\ii}+ip_{\jj}}\rb
\lb V^{\ii\jj}_{\tilde 2}-V^{\ii\jj}_{\tilde 3}\rb
\right]\;.
\eeq
Using this kind of expression with the explicit form of the potentials \eq{Vt} one  gets
\begin{multline}
\prod_{j=1}^{K_2}\frac{u_{1,k}-u_{2,j}+\frac{i}{2}}{u_{1,k}-u_{2,j}-\frac{i}{2}}
\prod_{j=1}^{K_4}\frac{1-1/(x_{1,k}x^+_{4,j})}{1-1/(x_{1,k}x^-_{4,j})}=
\exp
\biggl[\int_{-1}^{+1}\Bigl( r(1/x_{1,k},z)({\cal N}_{\hat 2 *}+{\cal N}_{\tilde 2 *})
\\  -
r(x_{1,k},z)({\cal N}_{\hat 1 *}+{\cal N}_{\tilde 1 *})
-u(x_{1,k},z){\cal N}_{*}-q(x_{1,k},z){\cal N}_{*}\Bigr) dz\biggr]\;,
\end{multline}
\begin{multline}
-\prod_{j=1}^{K_2}\frac{u_{2,k}-u_{2,j}-i}{u_{2,k}-u_{2,j}+i}
\prod_{j=1}^{K_1}\frac{u_{2,k}-u_{1,j}+\frac{i}{2}}{u_{2,k}-u_{1,j}-\frac{i}{2}}
\prod_{j=1}^{K_3}\frac{u_{1,k}-u_{3,j}+\frac{i}{2}}{u_{1,k}-u_{3,j}-\frac{i}{2}}=\\
\exp\left[\int_{-1}^{+1} (r(x_{2,k},z)+r(1/x_{2,k},z))({\cal N}_{\tilde 1 *}-{\cal N}_{\tilde 2 *}) dz\right]
\;,
\end{multline}
\begin{multline}
\prod_{j=1}^{K_2}\frac{u_{3,k}-u_{2,j}+\frac{i}{2}}{u_{3,k}-u_{2,j}-\frac{i}{2}}
\prod_{j=1}^{K_4}\frac{x_{3,k}-x^+_{4,j}}{x_{3,k}-x^-_{4,j}}=
 \exp
\biggl[\int_{-1}^{+1}\Bigl( r(x_{3,k},z)({\cal N}_{\hat 2 *}+{\cal N}_{\tilde 2 *})
\\ -
r(1/x_{3,k},z)({\cal N}_{\hat 1 *}+{\cal N}_{\tilde 1 *})+u(x_{3,k},z){\cal N}_{*}+q(x_{3,k},z){\cal N}_{*}\Bigr) dz\biggr]\;,
\end{multline}
\begin{multline}
-\lb\frac{x_{4,k}^{+}}{x_{4,k}^{-}}\rb^{L}\prod_{j=1}^{K_4}\sigma_{}^{-2}(u_{4,k},u_{4,j})
\frac{(1-1/(x_{4,k}^-x_{4,j}^+))(x_{4,k}^+-x_{4,j}^-)}{(1-1/(x_{4,k}^+x_{4,j}^-))(x_{4,k}^--x_{4,j}^+)}
\times
\\  \prod_{j=1}^{K_1}\frac{1-1/(x_{4,k}^-x_{1,j})}{1-1/(x_{4,k}^+x_{1,j})}
\prod_{j=1}^{K_3}\frac{x_{4,k}^--x_{3,j}}{x_{4,k}^+-x_{3,j}}
\prod_{j=1}^{K_5}\frac{x_{4,k}^--x_{5,j}}{x_{4,k}^+-x_{5,j}}
\prod_{j=1}^{K_7}\frac{1-1/(x_{4,k}^-x_{7,j})}{1-1/(x_{4,k}^+x_{7,j})}= \\
\exp\left[\int_{-1}^{+1}\lb r(1/x_{4,k},z)({\cal N}_{\hat 1 *}+{\cal N}_{*\hat 4})-r(x_{4,k},z)({\cal N}_{\hat 2*}+{\cal N}_{*\hat 3})-2u(x_{4,k},z){\cal N}_{*}\rb dz\right]
\end{multline}
where we  use the following kernels
\begin{align}
\nn & r(y,z)=\frac{y^2}{y^2-1}\frac{\partial_z}{2\pi g}\frac{1}{y-x(z)}\;\;,\qquad
u(y,z)=\frac{y}{y^2-1}\frac{\partial_z}{2\pi g}\frac{1}{x^2(z)-1}
\;\;, \\
 & q(y,z)= \frac{1}{y^2-1}\frac{\partial_z}{2\pi g}\frac{x(z)}{x^2(z)-1}\; \nn
\end{align}
and ${\cal N}$'s are some combinations of the quasi-momenta defined in \eq{eqN}.
The equations for $u_{5,k},u_{6,k},u_{7,k}$ could be easily written down following the obvious pattern of
modifying phases.
Together with the corrected equation for the energy
\beq
E=\sum_{j=1}^{K_4}\lb1+\frac{2ig}{x_{4,j}^+}-\frac{2ig}{x_{4,j}^-}\rb-\int_{-1}^{1}\frac{dz}{2\pi }
\partial_z \frac{z}{\sqrt{1-z^2}}
{\cal N}_*
\eeq
these equations define the  quasiclassical energy including one-loop contributions with all wrapping corrections included\footnote{In the derivation of these
equations usually one requires the filling fractions $K_a/L$ to be sufficiently small
and also the modifying phases are not too large. The  case of arbitrary filling fractions could be described  by an appropriate analytic continuation.}.

\end{document}